\documentclass[aps,onecolumn,pre]{revtex4}
\usepackage{latexsym}
\usepackage{graphicx}
\usepackage{amsfonts}
\usepackage{amssymb}
\usepackage{amsmath}
\usepackage[toc,page]{appendix}
\usepackage{float}
\usepackage{centernot}

\begin{document}

\title{\bf Wave Propagation in a Strongly Nonlinear Locally Resonant Granular Crystal}

\author{K. Vorotnikov}
\thanks{Corresponding author}
\email{kirill.vorotnikov@inria.fr}
\affiliation{Faculty of Mechanical Engineering, 
Technion Israel Institute of Technology, 
Technion City, Haifa 32000, Israel}

\author{Y. Starosvetsky}
\affiliation{Faculty of Mechanical Engineering, 
Technion Israel Institute of Technology, 
Technion City, Haifa 32000, Israel}

\author{G. Theocharis}
\affiliation{Laboratoire d'Acoustique de l'Universit\'{e} du Maine, UMR-CNRS 6613, Av. Olivier Messiaen, Le Mans 72000, France}

\author{P.G. Kevrekidis}
\affiliation{Department of Mathematics and Statistics, University of Massachusetts,Amherst MA 01003-4515, USA}

\begin{abstract}
In this work, we study the wave propagation in a recently proposed acoustic structure, the locally resonant granular crystal. This structure is composed of a one-dimensional granular crystal of hollow spherical particles in contact, containing linear resonators. The relevant model is presented
and examined through a combination of analytical approximations
(based on ODE and nonlinear map analysis) and of numerical results.
The generic dynamics of the system involves a degradation of the
well-known traveling pulse of the standard Hertzian chain of
elastic beads. Nevertheless, the present system is richer, in
that as the primary pulse decays, secondary ones emerge and eventually
interfere with it creating modulated wavetrains. Remarkably, upon
suitable choices of parameters, this interference ``distills''
a weakly nonlocal solitary wave (a ``nanopteron''). This motivates
the consideration of such nonlinear structures through a separate Fourier
space technique, whose results suggest the existence of
such entities not only with a single-side tail, but also with
periodic tails on both ends. These tails are found to oscillate
with the intrinsic oscillation frequency of the out-of-phase
motion between the outer hollow bead and its internal linear attachment.
\end{abstract}

\maketitle

\section{Introduction}

Dynamics of one-dimensional granular chains has attracted substantial 
interest from the researchers of quite different scientific areas~\cite{ref1,ref2,ref3,ref4,ref5,ref6,ref7,ref8,ref10,ref11,ref12,ref13,ref14,ref15,ref16,ref17,ref18,ref19,ref20,ref21,ref22,ref23,ref24,ref25}
due to their exciting 
dynamical properties. These chains support the formation of 
highly robust, strongly localized and genuinely traveling elastic stress waves.
The existence of traveling waves was originally proved in~\cite{ref7}
using the variational approach of~\cite{ref9}, yet no information
was given on their profile. Their single pulse character (in the strain
variables) was rigorously shown in~\cite{atanas}, following the
approach of~\cite{english}, and the doubly exponential character of
their spatial decay in the absence of precompression was established.
Earlier work on the basis of long wavelength approximations and
numerical computations had conjectured that the waves were genuinely
compact (spanning a finite number of elements)~\cite{ref1}.

Recent 
studies~\cite{ref10,ref11,ref12,ref13,ref14,ref15,ref16,ref17,ref18,ref19} 
in the area have been mainly concerned with the 
effect of various types of structural inhomogeneities induced in the granular chain. 
The latter leads to a modulation of the solitary waves, as well as to
new kinds of breathing modes produced either robustly~\cite{ref13,ref14,ref15,ref28,ref29}
or transiently~\cite{refjob}. Wave propagation in tapered and 
decorated granular chains has been extensively studied 
in~\cite{ref10,ref11,ref12} both 
analytically and numerically. The approximations developed in these 
works for the estimation of the maximal pulse velocity recorded on each one 
of the granules along with its propagation through such inhomogeneous
granular chains have demonstrated a good correspondence of the analytic 
predictions with the numerical simulations. Additional experimental, 
computational and analytical studies were devoted to the dynamics of the 
periodic granular chains (e.g. diatomic chains, granular 
containers, etc.) under various conditions of initial pre-compression 
\cite{ref13,ref14,ref15,ref16,ref17,ref18,ref19}. 
Dynamics of primary pulses in the non-compressed granular chain 
perturbed by a weak dissipation has been considered 
in~\cite{ref20,ref21,ref22,ref23,ref24,ref25}. These, in turn, 
shed light on the evolution of the primary pulses in the 
dissipative, 1D granular media and provide some qualitative theoretical 
(in some cases in connection with experiments~\cite{ref24})
estimations for modeling the dissipation in the chain as well as depicting 
the rate of decay of the primary pulse. 
A systematic theoretical attempt to capture the 
(decaying) evolution of a primary pulse in the granular chain subject to 
on-site perturbation has been provided in the extensive study of~\cite{ref26}. 

In the present paper we study a novel acoustic structure which has been recently considered in some experimental  and theoretical studies~\cite{LRGC,ref41,ref42},
the locally resonant granular crystal. The fundamental unit cell of these periodic systems is made of an outer mass (hollow spherical shell and an inner mass connected by a linear spring).
Our principal aim is to examine the dynamics of this novel
class of chains, as regards their ability to propagate 
travelling waves in the presence of these internal 
resonators. What we generally observe in this setting
is a decay of the principal pulse (in the strain variables),
due to its coupling to the internal variables.
The rate of such decay of a primary pulse can be fully captured 
analytically. Perhaps even more importantly,
 we show 
that,
depending on the parameters of the internal resonator (i.e. coupling stiffness 
and mass), the response can range from the above mentioned (continuous)
decay, to the formation of 
a  genuinely travelling primary pulse. The latter scenario is examined
in close detail and is identified as a case example of a ``nanopteron''
solution, whose tail carries an oscillation with the intrinsic 
linear frequency of the system (of the relative motion of the outer
and inner mass). To the best of our knowledge, 
and although such weakly nonlocal solitary waves have been studied
extensively in a series of examples in physical sciences and 
engineering~\cite{boyd},
this is only the second 
example of the reporting of such a nanopteronic solution 
in granular systems  (and the first where such waves are obtained as exact
solutions). The potential observation of such solutions 
in granular systems has been earlier suggested
based on the numerical observations of~\cite{ref29}, while this 
possibility has been proposed theoretically a decade ago
for FPU lattices in the work of~\cite{iooss}.
This motivates us to further examine the
problem using the methodology of~\cite{english}. As a result of
this study we illustrate that it is possible in fact to produce
nanoptera with oscillating tails on both ends of the principal
pulse. These results pave the way for a previously unexplored
class of solutions in these ``mass-in-mass'' systems and render
them especially intriguing candidates for experimental
investigations. 

Our presentation is structured as follows. In Section II,
we offer the basic setup and the corresponding theoretical model.
In Section III, we provide the analytical approach that
captures the typical (and systematic) decay of a primary
pulse in the ``mass-in-mass'' (hereafter abbreviated as MiM)
system. In Section IV, we test these predictions numerically,
obtaining good agreement with the analytical predictions, but
also shedding light on how a nanopteronic solution can be
seen to spontaneously emerge. In Section V, we propose a different
analytical-numerical approach for capturing such solutions
and offer a proof-of-principle confirmation of 
their existence (with tails on both sides of
the principal pulse). Finally, in Section VI, we summarize
our results and present our conclusions, as well as a number
of directions for future study. Lastly, in the appendix
we provide some technical details about the form of the traveling
wave in the homogeneous granular chain that are used in our theoretical
approach for capturing the decay of the primary pulse in the MiM
setting.

\section{Physical Model}

\begin{figure}[hbtp]
\centerline{\includegraphics[width=8.cm]{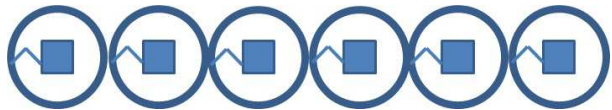}}
\caption{Scheme of the model under consideration}
\label{refig1}
\end{figure}

In the present study we consider the uncompressed, one-dimensional, locally resonant granular 
crystal composed of hollow elastic spheres in contact, containing linear resonators, as this is illustrated in Figure~\ref{refig1}. According to ~\cite{Hollow}, the contact interaction of two hollow spheres depends strongly on the thickness of the spherical shells. However, for relatively thick spherical shells, the interaction contact follows the Hertzian contact law~\cite{ref1}. 

The governing equations of motion can then be written as follows,
\begin{equation}
\label{eq1}
\begin{array}{l}
 M_{i} \frac{d^{2}U_{i} }{dt^{2}}=\left( {\frac{4}{3}} \right)E^{\ast }\sqrt 
R \left[ {\left( {U_{i-1} -U_{i} } \right)_{+}^{3 \mathord{\left/ {\vphantom 
{3 2}} \right. \kern-\nulldelimiterspace} 2} -\left( {U_{i} -U_{i+1} } 
\right)_{+}^{3 \mathord{\left/ {\vphantom {3 2}} \right. 
\kern-\nulldelimiterspace} 2} } \right]+k\left( {u_{i} -U_{i} } 
\right),\mbox{\thinspace }\forall i,\mbox{\thinspace }i\in N \\ 
 m_{i} \frac{d^{2}u_{i} }{dt^{2}}=-k\left( {u_{i} -U_{i} } \right). \\ 
 \end{array}
\end{equation}
Here $U_{i} $ is the displacement of the $i^{\rm th}$ sphere, while 
$u_{i} $ is the 
displacement of the small mass, linearly coupled
attachment inside the $i^{\rm th}$ sphere, $r_{i} 
$ is the radius of the sphere, $M_{i} $  is the mass of the sphere; and 
$E^{\ast }=E \mathord{\left/ {\vphantom {E {2\left( {1-\mu^{2}} \right)}}} 
\right. \kern-\nulldelimiterspace} {2\left( {1-\mu^{2}} \right)}$~; $E$ is 
the elastic (Young's) 
modulus and $\mu $  is the Poisson's ratio of the sphere. We 
note that the interaction force between the neighboring elements is given 
by $F=\left( {4 \mathord{\left/ {\vphantom {4 3}} \right. 
\kern-\nulldelimiterspace} 3} \right)E^{\ast }\sqrt{\frac{R_i}{2}} \Delta^{3 
\mathord{\left/ {\vphantom {3 2}} \right. \kern-\nulldelimiterspace} 2}$, 
where $r_i$ is bead radius (assumed implicitly to be uniform in the
above expression i.e., independent of $i$)
and $\Delta $  is their relative displacement. Moreover, 
the $\left( + \right)$  subscripts in (\ref{eq1}) and (\ref{eq3}) indicate that 
only non-negative values of the expressions in parentheses are considered,
i.e., the interaction is tensionless. We should also mention in passing
that a mathematically similar system with (isolated) 
external resonators (a so-called mass-with-mass setting)
has recently been proposed in~\cite{annavain}. It is noteworthy that though the mathematical equivalence of these two settings, their experimental realizations  are  quite different.

The system nondimensionalization is performed as follows
\begin{equation}
\label{eq2}
X_{i} =\frac{U_{i} }{r_{i} };\mbox{\thinspace \thinspace }x_{i} =\frac{u_{i} 
}{r_{i} };\mbox{\thinspace \thinspace }\tau =\left[ {\frac{E^{\ast }}{\sqrt 
2 \pi r_{i}^{2} \rho }} \right]^{1 \mathord{\left/ {\vphantom {1 2}} \right. 
\kern-\nulldelimiterspace} 2}t;\mbox{\thinspace \thinspace 
}\widetilde{\kappa }=\frac{3\sqrt 2 k}{4R_{i} E^{\ast }};\mbox{\thinspace 
\thinspace }\nu =\frac{m_{i} }{M_{i} }=\frac{m}{M}.
\end{equation}
It is important to note that in the present study we assume 
(with one caveat to be explained below) that the outer and inner masses 
are uniform all through the chain (i.e. $r_{i} =r,M_{i} =M,m_{i} =m)$.

Substituting (\ref{eq2}) into (\ref{eq1}) we end up with the following, non-dimensional 
set of equations governing the system dynamics
\begin{equation}
\label{eq3}
\begin{array}{l}
 X_{i,\tau \tau } =\left[ {\left( {X_{i-1} -X_{i} } \right)_{+}^{3 
\mathord{\left/ {\vphantom {3 2}} \right. \kern-\nulldelimiterspace} 2} 
-\left( {X_{i} -X_{i+1} } \right)_{+}^{3 \mathord{\left/ {\vphantom {3 2}} 
\right. \kern-\nulldelimiterspace} 2} } \right]+\widetilde{\kappa }\left( 
{x_{i} -X_{i} } \right)\mbox{\thinspace \thinspace \thinspace \thinspace } 
\\ 
 \nu x_{i,\tau \tau } =-\widetilde{\kappa }\left( {x_{i} -X_{i} } \right). \\ 
 \end{array}
\end{equation}
To make the further analysis of (\ref{eq3}) somewhat simpler it is convenient to 
introduce the coordinates of relative displacements (i.e.,
strains) for both outer and inner masses,
\begin{equation}
\label{eq4}
\Delta_{i} =X_{i} -X_{i+1} 
\end{equation}
\begin{equation}
\label{eq5}
d_{i} =x_{i} -x_{i+1}. 
\end{equation}
Substituting (\ref{eq4}) and (\ref{eq5}) into (\ref{eq3}) we obtain the following set of 
equations,
\begin{equation}
\label{eq6}
\begin{array}{l}
 \Delta_{i,\tau \tau } =\Delta_{\left( {i-1} \right),+}^{3 
\mathord{\left/ {\vphantom {3 2}} \right. \kern-\nulldelimiterspace} 2} 
-2\Delta_{\left( i \right),+}^{3 \mathord{\left/ {\vphantom {3 2}} 
\right. \kern-\nulldelimiterspace} 2} +\Delta_{\left( {i+1} \right), 
+}^{3 \mathord{\left/ {\vphantom {3 2}} \right. \kern-\nulldelimiterspace} 
2} +\widetilde{\kappa }\left( {d_{i} -\Delta_{i} } \right) \\ 
 \nu d_{i,\tau \tau } =-\widetilde{\kappa }\left( {d_{i} -\Delta_{i} } 
\right). \\ 
 \end{array}
\end{equation}
The goal of the present study is to examine the wave propagation along the contacts of the outer masses (which
contribute to highly nonlinear dynamics) in the presence of the internal mass attachments. We assume that the
coupling between the internal mass and the outer sphere is linear and weak such that $\widetilde{\kappa }$ treated as a small system
parameter $\widetilde{\kappa }=\varepsilon $. We anticipate that as the primary pulse propagates down the chain, it is possible for it to ``transfer"
energy to the internal, linear attachments, storing it in the form of potential energy and thus depriving the original
pulse from its initial kinetic energy. This, in turn, is expected to yield a decay of the amplitude of the primary pulse,
which we now consider in three distinct asymptotic limits, namely: (1) $\nu \ll 1$, (2) $\nu =O(1)$ and (3) $\nu \gg 1$.

Nevertheless, as we will show below, the phenomenology is not exhausted in the (very accurately captured through
our theoretical approach) pulse decay. In addition to it, we will illustrate the ability of the system to spontaneously
rearrange its propagating waveform into the shape of a nanopteron i.e., a weakly nonlocal solitary wave with a
non-vanishing tail that we subsequently study separately.

\section{Analytical Approximation of the Primary Pulse Transmission}

We start our analysis by noting that due to the linear coupling between outer and inner masses, one can readily
express the response of the relative displacement corresponding to inner masses by means of the relative displacement
of the heavy spheres. Thus, using the well-known theory of Green's functions one derives the solution for the relative
displacement of the attachments in the following form,

\begin{equation}
\label{eq7}
\begin{array}{l}
 d_{i} \left( t \right)=\varepsilon \nu^{-1}\int\limits_{-\infty }^t 
{g\left( {t\vert \tau } \right)\Delta_{i} \left( \tau \right)d\tau } 
=\omega_{0} \int\limits_{-\infty }^t {\sin \left( {\omega_{0} \left( 
{t-\tau } \right)} \right)\Delta_{i} \left( 
\tau \right)d\tau } \\ 
 g(t\vert \tau )=\omega_{0}^{-1}\sin \left( {\omega_{0} \left( {t-\tau } 
\right)} \right) \\ 
 \end{array}
\end{equation}
where $\omega_{0}^{2}=\varepsilon \nu^{-1}$, while $g\left( {t\vert \tau } 
\right)$ is a Green's function of the undamped linear oscillator. Substituting (\ref{eq7}) into 
the equation for the $\Delta_i(t)$, we obtain,
\begin{equation}
\label{eq8}
\begin{array}{l}
 \Delta_{i,tt} =\Delta_{\left( {i-1} \right),+}^{3 \mathord{\left/ 
{\vphantom {3 2}} \right. \kern-\nulldelimiterspace} 2} -2\Delta_{\left( i 
\right),+}^{3 \mathord{\left/ {\vphantom {3 2}} \right. 
\kern-\nulldelimiterspace} 2} +\Delta_{\left( {i+1} \right),+}^{3 
\mathord{\left/ {\vphantom {3 2}} \right. \kern-\nulldelimiterspace} 2} 
+\varepsilon Q(\Delta_{i} ,t) \\ 
 Q(\Delta_{i} ,t)=\left( {\omega_{0} \int\limits_{-\infty }^t {\sin \left( 
{\omega_{0} \left( {t-\tau } \right)} \right)\Delta_{i} \left( \tau \right)d\tau } -\Delta_{i}\left(\tau\right) } \right). \\ 
 \end{array}
\end{equation}
Further analysis of the system is based on the recently proposed analytical 
procedure of~\cite{ref26}, 
considering the evolution of the primary pulse along with its 
propagation through the granular chain subject to on-site perturbations. The 
main difference of the current system under consideration (\ref{eq10}) from the 
previously considered ones examined in~\cite{ref26} 
is in the functional form of the 
perturbation term $Q(\Delta_{i} ,t)$. To the best of the authors' knowledge, all 
the perturbed granular chains considered so far were autonomous,
contrary to what is the case herein.

An additional 
significant difficulty observed in (\ref{eq8}) is in 
the presence of the integral term 
in $Q(\Delta_{i} ,t)$. This means that in order to derive the analytical 
approximation for the spatial evolution of the primary pulse, one has to 
cope with an integro-differential equation (\ref{eq8}) rather than the 
ordinary 
differential equations analyzed previously. 

In this section we develop an analytical procedure depicting the evolution of the amplitude of the propagating
primary pulse interacting with the linear and weak, $\widetilde{\kappa }=\varepsilon $, local resonators. The proposed analytical approach is
based on the assumption that the functional form of the primary pulse 
can be approximated by the well-known
solitary wave solution, originally developed in~\cite{ref2} - see also ~\cite{ref1} for review{ derived for the unperturbed case of $\kappa = 0 $ which has a well known form of Nesterenko soliton derived in the continuum limit~\cite{ref1}. This takes the following form:

\begin{equation}
\label{eq9}
\Delta_i^s(\tau)=S^{s}\left(\tau-i\right)=
\begin{cases}
B \cos^4(\alpha(\tau-i)), &  \tau\in \left[{i-\frac{\pi}{2\alpha} , i+\frac{\pi}{2\alpha} } \right] \\
0, & \text{otherwise}
\end{cases}
\end{equation}
where $B$ and $\alpha$ corresponds to the amplitude and width of the normalized, solitary wave solution of unit phase
shift ($T$=1). These parameters have been calculated in such a way that the amplitude and the areas of exact and
approximate solitary wave profiles match.

In fact the parameters $\alpha$ and $B$  can be considered as universal (see the Appendix for details) and thus, the
general solitary wave solution propagating through the $i^{\rm th}$ contact element of the unperturbed chain can be written 
as

\begin{equation}
\label{eq10}
S_{i}(\tau)=AS^{s}\left(A^{1/4}\tau-i\right)
\end{equation}
where $A$ is the ratio between the amplitude of the arbitrary, solitary wave 
profile and the normalized on $S^{s}\left(0\right)$, and
$i$ is an index of the 
contact 
between the $i^{\rm th}$ and $(i+1)^{\rm th}$ elements of the 
granular chain.

\subsection{\bf Limit of small mass attachments $\nu 
\ll 1$}
Let us start with the analysis of a primary pulse transmission assuming the 
first asymptotic limit, $\tilde{\kappa}=\epsilon$. In this limit, $\omega_{0}$ is of $O(1)$.

Following the principal steps of~\cite{ref26,ref40}, for each contact in the chain we assign the local, semi-infinite 
time frame $t_{i}=t-T_{si}$, such that $t_{i}=0$ 
corresponds to the exact time point when the $i^{\rm th}$ contact 
reaches the first peak of the primary pulse of the propagating disturbance.
We assume that for the considered time interval $t_{i}\in \left( 
{-\infty ,0 } \right]$, the primary pulse response recorded on the 
$\left( {i-1} \right)^{th},i^{th},\left( {i+1} \right)^{th}$ contacts exhibits a solitary-like behavior and thus can be 
approximated as, $\Delta_{p} \left( t \right)=A_{p} \tilde{{S}}\left( 
{A_{p}^{1/4}t_{i}+i-p} \right)+o(\varepsilon )$, where $p=i-1, i, i+1$ , $T_{si} 
=T_{s(i-1)}+A_{i-1}^{-1/4}+\psi_{0}$ and $\psi_0$ controls the initial phase. Here, 
we set $\psi_0=0$. For the variable
$\tilde{S}$, we will again use the subscript $i$ to denote its value
at the $i^{\rm th}$ spatial site.

Next, plugging the above {\it ansatz}
into the integro-differential equation 
(\ref{eq8}) we obtain,
\begin{equation}
\label{eq11}
B^{-1 \mathord{\left/ {\vphantom {3 2}} \right. 
\kern-\nulldelimiterspace} 2}A_{i}^{3 \mathord{\left/ {\vphantom {3 2}} \right. 
\kern-\nulldelimiterspace} 2} \tilde{{{S}''}}_{i} =A_{i-1}^{3 
\mathord{\left/ {\vphantom {3 2}} \right. \kern-\nulldelimiterspace} 2} 
\widetilde{S}_{i-i}^{3 \mathord{\left/ {\vphantom {3 2}} \right. 
\kern-\nulldelimiterspace} 2} -2A_{i}^{3 \mathord{\left/ {\vphantom {3 2}} 
\right. \kern-\nulldelimiterspace} 2} \widetilde{S}_{i}^{3 \mathord{\left/ 
{\vphantom {3 2}} \right. \kern-\nulldelimiterspace} 2} +A_{i+1}^{3 
\mathord{\left/ {\vphantom {3 2}} \right. \kern-\nulldelimiterspace} 2} 
\widetilde{S}_{i+1}^{3 \mathord{\left/ {\vphantom {3 2}} \right. 
\kern-\nulldelimiterspace} 2} +B^{-1 \mathord{\left/ {\vphantom {3 2}} \right. 
\kern-\nulldelimiterspace} 2}\varepsilon A_{i} \left( {\omega_{0} 
\int\limits_{-\infty }^{t_{i}} {\sin \left( {\omega_{0} \left( {t_{i}-\tau } \right)} 
\right)\widetilde{S}_{i}\left( {A_{i}^{1 \mathord{\left/ {\vphantom {1 4}} 
\right. \kern-\nulldelimiterspace} 4} \tau} \right)d\tau } 
-\widetilde{S}_{i} } \right)+o(\varepsilon ).
\end{equation}
Here the notion $\tilde{S_{p}}=\tilde{S}\left(A_{p}^{1/4}t_{i}+i-p\right)=\begin{cases}
\cos^4(\alpha(A_{p}^{1/4}t_{i}+i-p)), &  t_{i}\in \left[{-\frac{\pi}{2\alpha A_{p}^{1/4}} , 0} \right] \\
0, & t_{i}\in \left[{-\infty , -\frac{\pi}{2\alpha A_{p}^{1/4}} } \right]
\end{cases} $, where $p=i-1,i,i+1$, has been used for the sake of brevity. It is important to note that (\ref{eq11}) is a valid approximation for a solitary 
like behavior in the time interval $(-\infty, 0)$.

The above ansatz has accounted for the time dependence in the problem
and has, in turn converted it into a {\it discrete} problem for
identifying the elements of the sequence $\{A_i\}$
which essentially depict the evolution of the primary pulse along with its 
propagation 
through the perturbed granular chain using an iterative procedure.

The next step in the approximation is a direct integration of (\ref{eq11}) in the local time interval $(-\infty,0]$,
leading to the following expression, 
\begin{equation}
\label{eq12}
\begin{array}{l}
 0=A_{i-1}^{3/2} \int\limits_{-\infty }^{0} {\tilde{S}^{3/2}\left(A_{i-1}^{1/4}t_{i}+1\right) 
dt_{i}} -2A_{i}^{3/2} \int\limits_{-\infty }^{0} {\tilde{S}^{3/2}\left(A_{i-1}^{1/4}t_{i}\right) 
dt_{i}} +A_{i+1}^{3/2} \int\limits_{-\infty }^{0} {\tilde{S}^{3/2}\left(A_{i-1}^{1/4}t_{i}-1\right) 
dt_{i}} \\ 
 \mbox{\thinspace \thinspace \thinspace \thinspace \thinspace \thinspace 
}+B^{-1 \mathord{\left/ {\vphantom {3 2}} \right. 
\kern-\nulldelimiterspace} 2}\varepsilon A_{i} \left[ {\omega_{0} \int\limits_{-\infty }^{0 } 
{\left( {\int\limits_{-\infty }^{t_{i}} {\sin \left( {\omega_{0} \left( {t_{i}-\tau } 
\right)} \right)\ \tilde{S}\left(A_{i}^{1/4}\tau\right)d\tau } } 
\right)dt_{i}} -\int\limits_{-\infty }^{0} {\tilde{S}\left(A_{i}^{1/4}t_{i}\right)dt_{i}} } \right]+o(\varepsilon ) \\ 
 \mbox{\thinspace \thinspace \thinspace \thinspace \thinspace \thinspace 
\thinspace \thinspace \thinspace \thinspace \thinspace \thinspace \thinspace 
\thinspace \thinspace \thinspace \thinspace \thinspace \thinspace \thinspace 
\thinspace \thinspace \thinspace \thinspace \thinspace \thinspace \thinspace 
\thinspace \thinspace \thinspace \thinspace \thinspace \thinspace \thinspace 
\thinspace \thinspace \thinspace \thinspace \thinspace \thinspace \thinspace 
\thinspace \thinspace \thinspace \thinspace \thinspace \thinspace \thinspace 
\thinspace \thinspace \thinspace \thinspace \thinspace \thinspace \thinspace 
\thinspace \thinspace \thinspace \thinspace \thinspace \thinspace \thinspace 
\thinspace \thinspace \thinspace \thinspace \thinspace \thinspace \thinspace 
\thinspace \thinspace \thinspace } \\ 
 t_{i}\in \left( {-\infty ,0}\right]. \\ 
 \end{array}
\end{equation}
It is important to note that the inertia term of (\ref{eq11}) vanishes in (\ref{eq12}) 
right after the integration in the specified time interval. This is 
precisely the reason for the choice of 
$T_{Si} $ as the time of the first peak of the 
primary pulse where the first derivative vanishes $\left(i.e., {\dot{{\Delta 
}}_{i} \left( t_{i} \right)\vert_{t_{i}=0} =0} \right)$.

By a proper scaling of time in the first, second, third and fifth terms of (\ref{eq12}) (i.e. $t_{i} \rightarrow A_{i-1}t_{i}$ (first term),$t_{i}\rightarrow A_{i}t_{i}$,(second and fifth terms), $t_{i}\rightarrow A_{i+1}t_{i}$ (third term))  one obtains the following 
simplification,
\begin{equation}
\label{eq13}
0=A_{i-1}^{5 \mathord{\left/ {\vphantom {5 4}} \right. 
\kern-\nulldelimiterspace} 4} f_{1} -2A_{i}^{5 \mathord{\left/ {\vphantom {5 
4}} \right. \kern-\nulldelimiterspace} 4} f_{2} +A_{i+1}^{5 \mathord{\left/ 
{\vphantom {5 4}} \right. \kern-\nulldelimiterspace} 4} f_{3} +B^{-1 \mathord{\left/ {\vphantom {3 2}} \right. 
\kern-\nulldelimiterspace} 2}\varepsilon 
\left( {A_{i} \omega_{0} \int\limits_{-\infty }^{0} {\left( 
{\int\limits_{-\infty }^{t_{i}} {\sin \left( {\omega_{0} \left( {t_{i}-\tau } 
\right)} \right)\widetilde{S}\left( {A_{i}^{1 \mathord{\left/ {\vphantom {1 
4}} \right. \kern-\nulldelimiterspace} 4} \tau} \right)d\tau } } 
\right)dt_{i}} -A_{i}^{3 \mathord{\left/ {\vphantom {3 4}} \right. 
\kern-\nulldelimiterspace} 4} g_{1000} } \right)+o(\varepsilon )
\end{equation}
where $f_{1} =\int\limits_{-\infty }^0 {\widetilde{S}^{3 \mathord{\left/ 
{\vphantom {3 2}} \right. \kern-\nulldelimiterspace} 2}\left( {t_{i}+1} 
\right)dt_{i}} =1.4568,\mbox{\thinspace }f_{2} =\int\limits_{-\infty }^0 
{\widetilde{S}^{3 \mathord{\left/ {\vphantom {3 2}} \right. 
\kern-\nulldelimiterspace} 2}\left( t_{i} \right)dt_{i}} =0.7615,\mbox{\thinspace 
}f_{3} =\int\limits_{-\infty }^0 {\widetilde{S}^{3 \mathord{\left/ 
{\vphantom {3 2}} \right. \kern-\nulldelimiterspace} 2}\left( {t_{i}-1} 
\right)dt_{i}} =0.0663$
and
$g_{1000} =\int\limits_{-\infty }^0 {\widetilde{S}\left( t_{i} \right)dt_{i}} 
=0.9138$. These integrations have been carried out by utilizing the
approximate form of the travelling solitary wave solution; see the
Appendix for details.

Here we make a note that the developed analytical procedure applied 
on (\ref{eq12}) ultimately connects $A_{i-1}$ with $A_i$ and $A_{i+1}$.
Thus, it is not sufficient for extracting the sequence of $A_i$'s
and necessitates additional assumptions. A reasonable one such is that the variation of the 
amplitude of the primary pulse profile propagating from contact to contact 
is bounded by the order of applied perturbation, namely

\begin{equation}
\label{eq14}
A_{i+1} =A_{i} +\varepsilon \Delta +o(\varepsilon ).
\end{equation}

Moreover, we identify an additional small parameter in (\ref{eq13}), 
\begin{equation}
\label{eq15}
\mu =
\raise0.7ex\hbox{${f_{3} }$} \!\mathord{\left/ {\vphantom {{f_{3} } 
{f_{1} }}}\right.\kern-\nulldelimiterspace}\!\lower0.7ex\hbox{${f_{1} 
}$} \ll 1,\mbox{\thinspace \thinspace }\mu \ll \raise0.7ex\hbox{${f_{2} }$} 
\!\mathord{\left/ {\vphantom {{f_{2} } {f_{1} 
}}}\right.\kern-\nulldelimiterspace}\!\lower0.7ex\hbox{${f_{1} }$}.
\end{equation}
Thus, maintaining the first term stemming from the right hand
side of~(\ref{eq14}) and neglecting terms $\propto \epsilon \mu$,
while preserving only terms $\propto \epsilon$ and $\propto \mu$, 
we have that: 

\begin{equation}
\label{eq16}
\begin{array}{l}
 0=A_{i-1}^{5/4}-2A_{i}^{5/4}\left( {\raise0.7ex\hbox{${f_{2} }$} 
\!\mathord{\left/ {\vphantom {{f_{2} } {f_{1} 
}}}\right.\kern-\nulldelimiterspace}\!\lower0.7ex\hbox{${f_{1} }$}} 
\right)+A_{i}^{5/4}\mu \\ 
+B^{-1 \mathord{\left/ {\vphantom {3 2}} \right. 
\kern-\nulldelimiterspace} 2}\frac{\varepsilon}{f_{1}} \left( {A_{i} \omega_{0} \int\limits_{-\infty }^{0} 
{\left( {\int\limits_{-\infty }^{t_{i}} {\sin \left( {\omega_{0} \left( {t_{i}-\tau } 
\right)} \right)\widetilde{S}\left( {A_{i}^{1 \mathord{\left/ {\vphantom {1 
4}} \right. \kern-\nulldelimiterspace} 4} \tau} \right)d\tau } } 
\right)dt_{i}} -A_{i}^{3 \mathord{\left/ {\vphantom {3 4}} \right. 
\kern-\nulldelimiterspace} 4} g_{1000} } \right)+O(\varepsilon \mu 
)+o(\varepsilon ). \\ 
 \end{array}
\end{equation}
 
At this level of approximation, (\ref{eq16}) defines a 
nonlinear, one-dimensional map ($A_{i-1} \to A_{i} )$, which can not
be explicitly solved. Equation (\ref{eq16}) is 
solved iteratively at each step determining the resulting amplitude of the 
primary pulse $\left\{ {A_{i} } \right\}$. 

The system (\ref{eq16}) can be rewritten in a more compact form
\begin{equation}
\label{eq17}
A_{i}^{5 \mathord{\left/ {\vphantom {5 4}} \right. 
\kern-\nulldelimiterspace} 4} \left( {-2\frac{f_{2}}{f_{1}} +\mu } \right)+B^{-1 \mathord{\left/ {\vphantom {3 2}} \right. 
\kern-\nulldelimiterspace} 2}\frac{\varepsilon}{f_{1}} 
\left( {A_{i} I\left( {\omega_{0} ,A_{i} } \right)-A_{i}^{3 \mathord{\left/ 
{\vphantom {3 4}} \right. \kern-\nulldelimiterspace} 4} g_{1000} } 
\right)+A_{i-1}^{5 \mathord{\left/ {\vphantom {5 4}} \right. 
\kern-\nulldelimiterspace} 4} =0
\end{equation}
where $I\left( {\omega_{0} ,A_{i} } \right)=\omega_{0} 
\int\limits_{-\infty }^{0} {\left( {\int\limits_{-\infty }^{t_{i}} {\sin 
\left( {\omega_{0} \left( {t_{i}-\tau } \right)} \right)\widetilde{S}\left( 
{A_{i}^{1 \mathord{\left/ {\vphantom {1 4}} \right. 
\kern-\nulldelimiterspace} 4} \tau} \right)d\tau } } \right)dt_{i}}=\omega_{0} A_{i}^{-1/2} \int\limits_{-\infty }^{0} {\left( {\int\limits_{-\infty }^{\xi} {\sin\left(\omega_{0} A_{i}^{-1/4}\left(\xi-\phi \right)\right)\widetilde{S}\left(\phi \right) d\phi} }\right)d\xi}. $

The direct integration of (\ref{eq17}), yields the explicit nonlinear map as following:

\begin{equation}
\label{eq18}
A_{i}^{5 \mathord{\left/ {\vphantom {5 4}} \right. 
\kern-\nulldelimiterspace} 4} \left( {-2\frac{f_{2}}{f_{1}} +\mu } \right)+B^{-1 \mathord{\left/ {\vphantom {3 2}} \right. 
\kern-\nulldelimiterspace} 2}\frac{\varepsilon}{f_{1}} 
\left( {A_{i}^{3 \mathord{\left/ {\vphantom {3 4}} \right. 
\kern-\nulldelimiterspace} 4} \left( \frac{3\pi}{16\alpha}-g_{1000} \right)-\frac{24\alpha^{4}A_{i}^{2} \sin\left(\frac{\omega_{0}\pi}{2\alpha A_{i}^{1/4}} \right)}{\omega_{0}\left(\omega_{0}^{4}-20\omega_{0}^{2}\alpha^{2} A_{i}^{1/2}+64\alpha^{4} A_{i} \right)} } 
\right)+A_{i-1}^{5 \mathord{\left/ {\vphantom {5 4}} \right. 
\kern-\nulldelimiterspace} 4} =0.
\end{equation}

Fortunately, for the case of uniformly perturbed chains (i.e., $\epsilon$ and $\omega_{0}$ remain constants all through the chain), the expression (\ref{eq18}) can be homogenized. 

Thus, performing the transformation $A_{i}^{5/4}=D_{i}$ leads to
\begin{equation}
\label{eq100}
0=f_{1}D_{i-1}-2f_{2} D_{i}+f_{3} D_{i+1}+\epsilon B^{-1/2} D_{i}^{3/5}\left[\left(\frac{3\pi}{16\alpha}-g_{1000} \right)-\frac{24\alpha^{4}D_{i} \sin\left(\frac{\omega_{0}\pi}{2\alpha D_{i}^{1/5}} \right)}{\omega_{0}\left(\omega_{0}^{4}-20\omega_{0}^{2}\alpha^{2} D_{i}^{2/5}+64\alpha^{4} D_{i}^{4/5} \right)} \right]. 
\end{equation}

A long-wave approximation then yields the following transformation:
\begin{equation}
\label{eq101}
D_{i} = D,\,\,\,\,  D_{i-1} = D - hD' + O\left(h^{2} \right),\,\,\,\,  D_{i+1} = D + hD' + O\left(h^{2} \right). 
\end{equation}

After the substitution, we obtain the ODE of the first order
\begin{equation}
\label{eq102}
hD' = \frac{\epsilon B^{-1/2} D^{3/5}}{f_{3}-f_{1}}\left[\frac{24\alpha^{4}D \sin\left(\frac{\omega_{0}\pi}{2\alpha D^{1/5}} \right)}{\omega_{0}\left(\omega_{0}^{4}-20\omega_{0}^{2}\alpha^{2} D^{2/5}+64\alpha^{4} D^{4/5} \right)} -\left(\frac{3\pi}{16\alpha}-g_{1000} \right)  \right].
\end{equation}

It is obvious that, in the normalized system under investigation $h = 1$, and thus, the system is rewritten as
\begin{equation}
\label{eq103}
D' = \frac{\epsilon B^{-1/2} D^{3/5}}{f_{3}-f_{1}}\left[\frac{24\alpha^{4}D \sin\left(\frac{\omega_{0}\pi}{2\alpha D^{1/5}} \right)}{\omega_{0}\left(\omega_{0}^{4}-20\omega_{0}^{2}\alpha^{2} D^{2/5}+64\alpha^{4} D^{4/5} \right)} -\left(\frac{3\pi}{16\alpha}-g_{1000} \right)  \right].
\end{equation}

It is worthwhile to note that after computing the drop in the response recorded on the contacts of the outer spheres, one can recover explicitly the primary response of the internal resonators. Thus preforming the integration of (\ref{eq7}) one arrives at the following form:
\begin{equation}
\label{eq108}
d_{i}(t_{i})=
\begin{cases}
0, &  t_{i}\in \left(-\infty ,-A_{i}^{-1/4} \frac{\pi}{2\alpha} \right] \\
C_{1}\cos^{4}\left(\alpha A_{i}^{1/4}t_{i}\right)+C_{2}\left(1-\cos\left(\omega_{0}t_{i}+\frac{\pi\omega_{0}}{2\alpha A_{i}^{1/4}}\right)\right)+C_{3}\cos^{2}\left(\alpha A_{i}^{1/4}t_{i}\right), & t_{i}\in \left[-A_{i}^{-1/4}\frac{\pi}{2\alpha}, A_{i}^{-1/4}\frac{\pi}{2\alpha}\right] \\
U\sin\left(\omega_{0}t_{i}\right)\sin\left(\frac{\pi\omega_{0}}{2\alpha A_{i}^{1/4}}\right), & t_{i}\in \left[A_{i}^{-1/4}\frac{\pi}{2\alpha}, +\infty \right) \\
\end{cases}
\end{equation}
where $C_{1}$=$\frac{-4A_{i}B\omega_{0}^{2}\left(\alpha^{2}A_{i}^{1/2}-\frac{\omega_{0}^{2}}{4}\right)}{K} $, $C_{2}$=$\frac{24\alpha^{4}A_{i}^{2}B}{K} $, $C_{3}$=$\frac{-12\alpha^{2}\omega_{0}^{2}A_{i}^{3/2}B}{K} $, $U$=$\frac{48A_{i}^{2}\alpha^{4}B}{K} $ and $K$=$\omega_{0}^{4}-20\omega_{0}^2\alpha^{2} A_{i}^{1/2}+64\alpha^{4} A_{i}$.
Nevertheless, it should be highlighted that the above obtained
quadrature solution provides a rather uncontrolled approximation,
as there is no small parameter for this expansion (notice that
$h$ has been set to unity). On the other hand, that is an approximation
that occasionally works unexpectedly well in the context of
granular crystals~\cite{ref1}.

\subsection{Comparable masses $\nu =O(1)$}

In the present subsection we analyze the evolution of the primary pulse in 
the case of the internal mass being comparable to that of the outer sphere. 
Though this case can be directly addressed by the already developed 
approximation (\ref{eq18}), the main purpose of the present 
subsection is 
to show that this approximation can be simplified even more in the case of 
$\nu =O(1)$.

Starting from the construction of the map, we have
\begin{equation}
\label{eq20}
A_{i}^{5 \mathord{\left/ {\vphantom {5 4}} \right. 
\kern-\nulldelimiterspace} 4} \left( {-2\frac{f_{2}}{f_{1}} +\mu } \right)+B^{-1 \mathord{\left/ {\vphantom {3 2}} \right. 
\kern-\nulldelimiterspace} 2}\frac{\varepsilon}{f_{1}} 
\left( {A_{i}^{3 \mathord{\left/ {\vphantom {3 4}} \right. 
\kern-\nulldelimiterspace} 4} \left( \frac{3\pi}{16\alpha}-g_{1000} \right)-\frac{24\alpha^{4}A_{i}^{2} \sin\left(\frac{\omega_{0}\pi}{2\alpha A_{i}^{1/4}} \right)}{\omega_{0}\left(\omega_{0}^{4}-20\omega_{0}^{2}\alpha^{2} A_{i}^{1/2}+64\alpha^{4} A_{i} \right)} } 
\right)+A_{i-1}^{5 \mathord{\left/ {\vphantom {5 4}} \right. 
\kern-\nulldelimiterspace} 4} =0.
\end{equation}

Accounting for  $\nu =O(1)$ and expanding the term containing
the $\sin$ in Taylor series (with respect to $\epsilon$) up to the first order, yields the following simplification of the nonlinear map (\ref{eq20}),

\begin{equation}
\label{eq21}
0=A_{i-1}^{5 \mathord{\left/ {\vphantom {5 4}} \right. 
\kern-\nulldelimiterspace} 4}-2A_{i}^{5 \mathord{\left/ {\vphantom {5 
4}} \right. \kern-\nulldelimiterspace} 4} \frac{f_{2}}{f_{1}} +A_{i}^{5 \mathord{\left/ 
{\vphantom {5 4}} \right. \kern-\nulldelimiterspace} 4} \mu -B^{-1/2}\frac{\varepsilon}{f_{1}} 
\left( {A_{i}^{3 \mathord{\left/ {\vphantom {3 4}} \right. 
\kern-\nulldelimiterspace} 4} g_{1000} } \right).
\end{equation}

As the local oscillator acts now purely on the outer shell of
our granular hollow spheres, 
 the evolution of the 
primary pulse becomes tantamount to that of  
the granular chain, mounted on a linear elastic foundation,
a case examined in~\cite{ref26}. Again, arguing exactly as above (i.e. (\ref{eq101})--(\ref{eq103})) the appropriate long wave approximation can be explicitly derived for the nonlinear map (\ref{eq21}). Here we would like to emphasize  that unlike the previous case, in the case of the heavy internal masses the expression for the long wave approximation depicting the spatial evolution of the primary pulse, takes the following explicit form:

\begin{equation}
\label{eq22}
A=\left\{ {A_{0}^{1/2}-\frac{2}{5}\frac{B^{-1/2}\varepsilon g_{1000} \xi }{f_{1} 
 }} \right\}^{2}
\end{equation}
where $A_{0}$ is the amplitude of the solitary wave pulse impinging on the 
perturbed part of the chain and $\xi$ is the continuum variable associated
with the site $i$ in the long wavelength limit developed in~\cite{ref26}.

\subsection{ Limit of the heavy internal masses $\nu 
\gg 1$}

In the present subsection we consider the case of $\nu \gg 1$. We again start from the 
rescaled system (\ref{eq3}).
Keeping in mind that $\varepsilon $ is a small parameter and accounting for a 
high mass mismatch ($\nu \gg 1)$, we can rewrite the system as
\begin{equation}
\label{eq24}
\begin{array}{l}
 X_{i,\tau \tau } =\left[ {\left( {X_{i-1} -X_{i} } \right)_{+}^{3 
\mathord{\left/ {\vphantom {3 2}} \right. \kern-\nulldelimiterspace} 2} 
-\left( {X_{i} -X_{i+1} } \right)_{+}^{3 \mathord{\left/ {\vphantom {3 2}} 
\right. \kern-\nulldelimiterspace} 2} } \right]+\varepsilon \left( {x_{i} 
-X_{i} } \right)\mbox{\thinspace \thinspace \thinspace \thinspace } \\ 
 x_{i,\tau \tau } =0 \\ 
 \end{array}
\end{equation}
by letting $\varepsilon /\nu \to 0$.  This, coupled with the
vanishing initial conditions on the internal oscillators, 
yields the immediate reduction,
\begin{equation}
\label{eq25}
X_{i,\tau \tau } =\left[ {\left( {X_{i-1} -X_{i} } \right)_{+}^{3 
\mathord{\left/ {\vphantom {3 2}} \right. \kern-\nulldelimiterspace} 2} 
-\left( {X_{i} -X_{i+1} } \right)_{+}^{3 \mathord{\left/ {\vphantom {3 2}} 
\right. \kern-\nulldelimiterspace} 2} } \right]-\varepsilon X_{i} 
\mbox{\thinspace \thinspace \thinspace \thinspace }
\end{equation}
which once again leads 
to the dynamics of a homogeneous granular chain mounted 
on an elastic foundation. 

In fact in the case of hard excitation, (i.e., for 
$X_{i} =O(1))$ the response of 
(\ref{eq25}) can be described by a nonlinear mapping procedure (\ref{eq21}) and (\ref{eq22}) 
depicting the modulation of solitary like pulses. However, for the case of 
the low amplitude excitations (i.e., for $X_{i} =O(\varepsilon^{2}))$ system 
(\ref{eq25}) can be immediately rescaled, ($X_{i} =\varepsilon^{2}\bar{{x}}_{i} 
,\bar{{t}}=\sqrt \varepsilon \tau )$ yielding
\begin{equation}
\label{eq26}
\bar{{x}}_{i}^{\prime \prime }+\bar{{x}}_{i} =\left[ {\left( 
{\bar{{x}}_{i-1} -\bar{{x}}_{i} } \right)_{+}^{3 \mathord{\left/ {\vphantom 
{3 2}} \right. \kern-\nulldelimiterspace} 2} -\left( {\bar{{x}}_{i} 
-\bar{{x}}_{i+1} } \right)_{+}^{3 \mathord{\left/ {\vphantom {3 2}} \right. 
\kern-\nulldelimiterspace} 2} } \right]\mbox{\thinspace \thinspace }.
\end{equation}
This system has been recently considered in~\cite{ref28,ref29}.
There, it was shown analytically and numerically that 
this system supports various periodic solutions, ranging from spatially 
periodic standing and traveling waves up to strongly localized ones (i.e. 
standing and moving breathers). 

\section{Numerical Simulations}

\subsection{Model Setup}
 We now consider detailed numerical simulations of the model. The goal is twofold, to examine the predictions of
the analytical approximations based on the nonlinear map approach of the previous section and to reveal additional
features of the emerging coherent structures. Regarding the
first goal, it is important to emphasize that the analytical
scheme of the nonlinear map, assumes that the propagating primary pulse has a waveform in the shape of a Nesterenko
soliton ~\cite{ref1,ref2}.
\begin{figure}[hbtp]
\centerline{\includegraphics[width=8.cm]{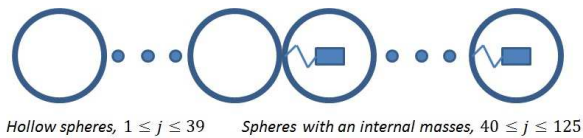}}
\caption{Scheme of the model which is used in numerical simulations.}
\label{refig2}
\end{figure}
Therefore, to assure that the primary pulse impinging on the perturbed part of the chain (the MiM
lattice) has already assumed a solitary waveform, we select the initial part of the chain to be composed of typically
39 uniform, hollow spheres (cf. Figure~\ref{refig2}) and we consider a velocity impulsive excitation.

 Thus, the dynamical model under consideration in the present numerical study reads: 

\begin{equation}
\label{eq29}
\begin{array}{l}
{X_{i,\tau \tau }} = (1 - {\delta _{1,i}})\left( {{X_{i - 1}} - {X_i}} \right)_ + ^{3/2} - \left( {{X_i} - {X_{i + 1}}} \right)_ + ^{3/2}{\rm{                               }}{\rm ,} \quad \quad 2 \le i < 41\\
{X_{i,\tau \tau }} = \left( {{X_{i - 1}} - {X_i}} \right)_ + ^{3/2} - (1 - {\delta _{i,125}})\left( {{X_i} - {X_{i + 1}}} \right)_ + ^{3/2} + \varepsilon \left( {{x_i} - {X_i}} \right){\rm{,      }} \quad 41 \le i \le 125\\
\nu {x_{i,\tau \tau }} =  - \varepsilon \left( {{x_i} - {X_i}} \right){\rm{\,                                                                    }}, 
\quad \quad 41 \le i \le 125\\
\end{array}
\end{equation}

where $\delta_{1,i}$ and $\delta_{i,125}$ denote Kronecker symbols, and

\begin{equation}
\label{291}
\begin{array}{l}

{X_{1,\tau }}\left( 0 \right) = {\widetilde V_0},{X_1}(0) = 0\\
{X_i}\left( 0 \right) = {X_{i,\tau }}\left( 0 \right) = 0, \quad \quad {\rm{  2}} \le i \le 125{\rm{ }}\\
{x_i}(0) = {x_{i,\tau }}(0) = 0{\rm{,    }} \quad \quad 41 \le i \le 125.
\end{array}
\end{equation}

From extensive simulations of this numerical configuration, we can distinguish the following three distinct cases:

Case (i): Weak coupling ($\varepsilon \ll 1$), 
light inner masses ($\nu = O(\varepsilon)$), initial excitation of the order of one ($\tilde{V}_{0} = O(1)$).

Case (ii): Weak coupling ($\varepsilon \ll 1$), comparable or heavier inner masses ($\nu = O(1)$ or $\nu \ge 1$), initial excitation of the order of one ($\tilde{V}_{0} = O(1)$).

Case (iii): Weak to moderate coupling ($\varepsilon = O(1)$), heavier inner masses($\nu \ge 1$), weak initial excitation ($\tilde{V}_{0} \ll 1$).

Intuitively, assuming the initial excitation to be of order $\tilde{V}_{0} = O(1)$ the first two asymptotic cases correspond
to the weak perturbation of the Nesterenko solitary wave. Clearly these two cases correspond to the hard
excitation, as the strength of the initial excitation is much higher than that of the coupling term $\varepsilon$.

In fact for the case of a hard excitation (cases (i) and (ii)) it is rather convenient to perform an additional rescaling of (\ref{eq29}). Thus introducing the new scales of time and displacements
\begin{equation}
\label{eq30}
\tilde \tau  = \left( {\tilde V_0^{1/5}} \right)\tau,\,\,\,\,\,{\tilde X_i} = \left( {\tilde V_0^{ - 4/5}} \right){X_i},\,\,\,\,\,{\tilde x_i} = \left( {\tilde V_0^{ - 4/5}} \right){x_i} 
\end{equation}
the rescaled system under consideration has a unit velocity kick
(absorbed in the rescaling) and a renormalized $\tilde \varepsilon  = \varepsilon \tilde{V_0}^{-2/5}$.

Accounting for the rescaling brought in (\ref{eq30}), the limiting cases can be slightly reformulated as:

Case (i): $\varepsilon \ll 1$, $\nu = O(\varepsilon)$.

Case (ii): $\varepsilon \ll 1$, $\nu = O(1)$ or $\nu \ge 1$.

Case (iii): $\varepsilon \gg 1$, $\nu \ge 1$.

As it was pointed out in several works studying the dynamics of solitary waves impinging on the interface of
the (mass mismatched) granular chains, in such cases 
there is a formation of a train of the transmitted and reflected pulses. In
fact, in the framework of our study for the cases (i) and (ii), we adopt a certain asymptotic assumption, under which
the perturbation induced by the local resonators is treated as a small parameter. This formally provides us with the
right to assume that the primary pulse only slightly deviates from the unperturbed solution on each site (Nesterenko
soliton). The third limiting case is rather different from cases (i) and (ii). In fact, unlike the first two cases which
correspond to the evolution of solitary like pulses in the perturbed chain, case (iii) concerns the formation of moving
breathers, previously studied in ~\cite{ref28,ref29}. In the rest of the paper all the numerical simulations corresponding to the asymptotic cases (i) and (ii) have been performed for the rescaled model, while
when these related to the case (iii) have been performed for original one 
of (\ref{eq29}).   

\subsection{Numerical Study of the Primary Pulse Transmission}

A typical result of the system's evolution
is presented in Figure~\ref{refig3} for 
{$\tilde \varepsilon =0.05$} and {$\mbox{\thinspace }\nu =0.128$}.
Here, we plot the evolution of the relative displacement 
for numerous sites throughout the chain. These reveal a primary pulse 
response of the system, forming as a result of 
the impulsive excitation supplied 
to the left end of the chain. As it is clear from the results of the figure, 
three main stages can be distinguished in the propagation of the
primary pulse; (1) the 
undisturbed propagation of a solitary wave through the first part of the 
chain free of local resonators (this propagation can be observed in 
Region 1) (2) the systematic drop in the amplitude of a single-hump 
primary pulse due to its interaction with the local resonators (Region 
2) (3) formation of the complex (`breathing') modulated, multi-humped 
pulse patterns 
(Region 3). In the currently examined limit of $\tilde{\varepsilon} \ll 1$
and of $\nu  = O(\tilde{\varepsilon})$, these findings can be considered
as rather generic.

\begin{figure}[hbtp]
\centerline{\includegraphics[width=0.90\textwidth]{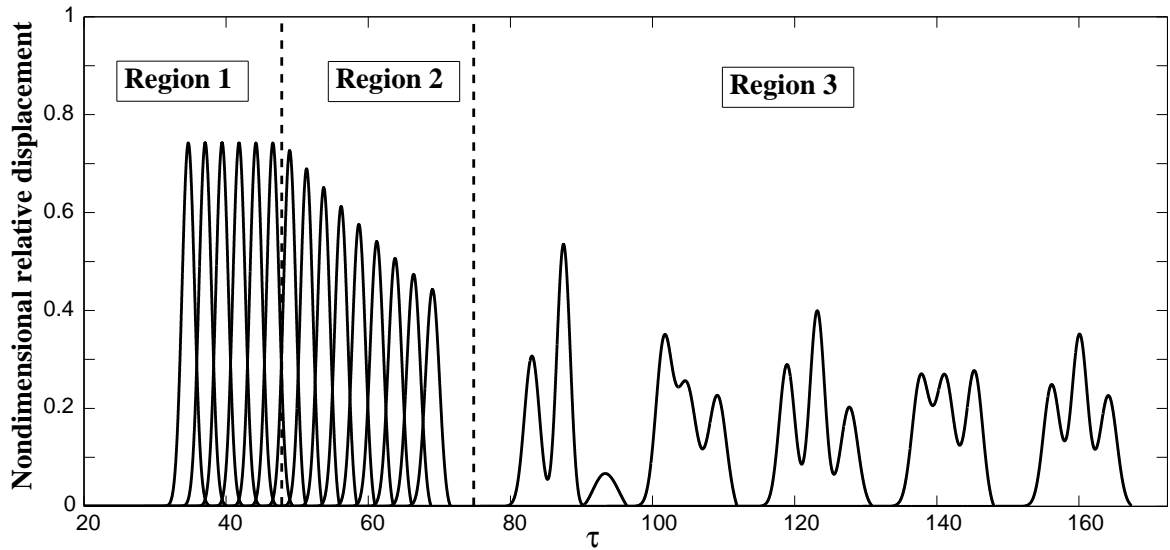}}
\caption{Time histories of 
relative displacements for a number of different contacts in the chain,
revealing
the pulses recorded at the 
different contacts of the outer spheres. Parameters of the linear resonators: 
$\tilde \varepsilon =0.05,\mbox{\thinspace }\nu =0.128$}
\label{refig3}
\end{figure}

A somewhat unexpected feature in the context of this model 
concerns the formation of the modulated wavetrains presented above
in Figure~\ref{refig3}, on which we now focus.
To better visualize their mechanism of formation, 
we illustrate the gradual evolution of a secondary 
hump emerging along the chain in Figure~\ref{refig4}, 
as the primary pulse decays. At the initial stage of the process 
the two humps (primary and secondary) 
are well separated, amplitude-wise. 
Alongside the decay of a single humped primary pulse, 
one observes the formation and gradual growth of the secondary 
(single-humped) pulse. On some specific contact of the chain the 
amplitude of the secondary hump exceeds the amplitude of the first one. 
Moreover, at some point the two humps (or more) merge into a modulated, 
multi-humped primary pulse (Region 3 of Figure~\ref{refig3}). 
This modulation of the 
primary pulse constitutes a qualitatively new regime which obviously differs 
from the initial un-attenuated propagation of the solitary pulse (of Region 1) 
as well as from the monotonic decay of a single humped primary pulse 
(of Region 2). 

\begin{figure}[hbtp]
\centerline{\includegraphics[width=0.90\textwidth]{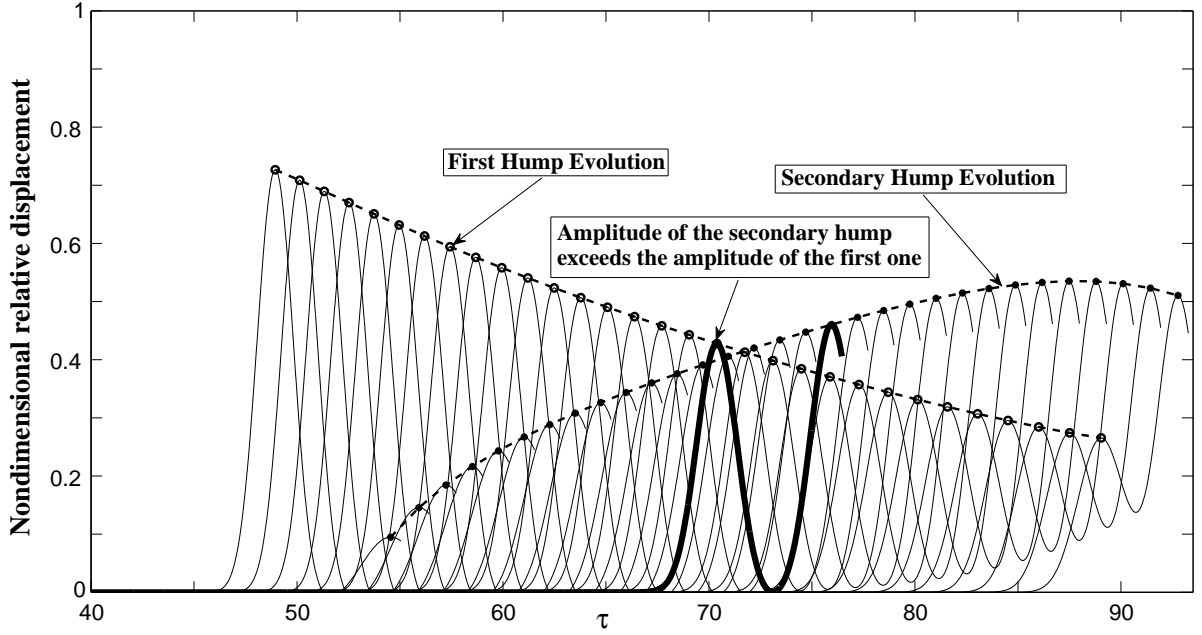}}
\caption{Time histories (corresponding to the numerical simulation) revealing the evolution of the primary and secondary 
humps leading to the formation of the multi-humped primary pulses recorded 
on the different contacts of the outer spheres. Parameters of the linear 
resonators: $\tilde \varepsilon =0.05,\mbox{\thinspace }\nu =0.128$. The 
evolution of the 
amplitudes of the primary and secondary humps is denoted by  
lines as a guide to the eye. }
\label{refig4}
\end{figure}

Clearly, the analytical procedure developed in the previous section 
can only be valid in the Region 2 where the waveform of the primary 
pulse profile exhibits a single hump decaying due to the
perturbation, yet preserving its 
solitary wave behavior, justifying the approximations
of our analytical approach which we now test in detail. 

\subsection{Numerical Verifications of the Analytical Model}

In this subsection we perform the detailed comparison between the analytical 
approximations of (\ref{eq18}) and (\ref{eq22}) predicting the evolution of the primary pulses (in Region 2) 
with the direct numerical simulations for the three main cases.

{\bf Case (i): }

In Figures \ref{refig8} and \ref{refig9},
 we plot the evolution of the primary pulses in the second 
region for two fixed values ($\nu =0.05$ and $0.25$; similar results
have been obtained for larger values e.g. for $\nu=0.55$) of the internal 
masses, gradually increasing the value of the stiffness coefficient. 
 
It is 
important to note that for the sake of better illustration of the global 
picture of a primary pulse transmission, in some figures we have also 
plotted the formation and evolution of the secondary humps. 

\begin{figure}[htbp]
\centerline{\includegraphics[width=1.05\textwidth]{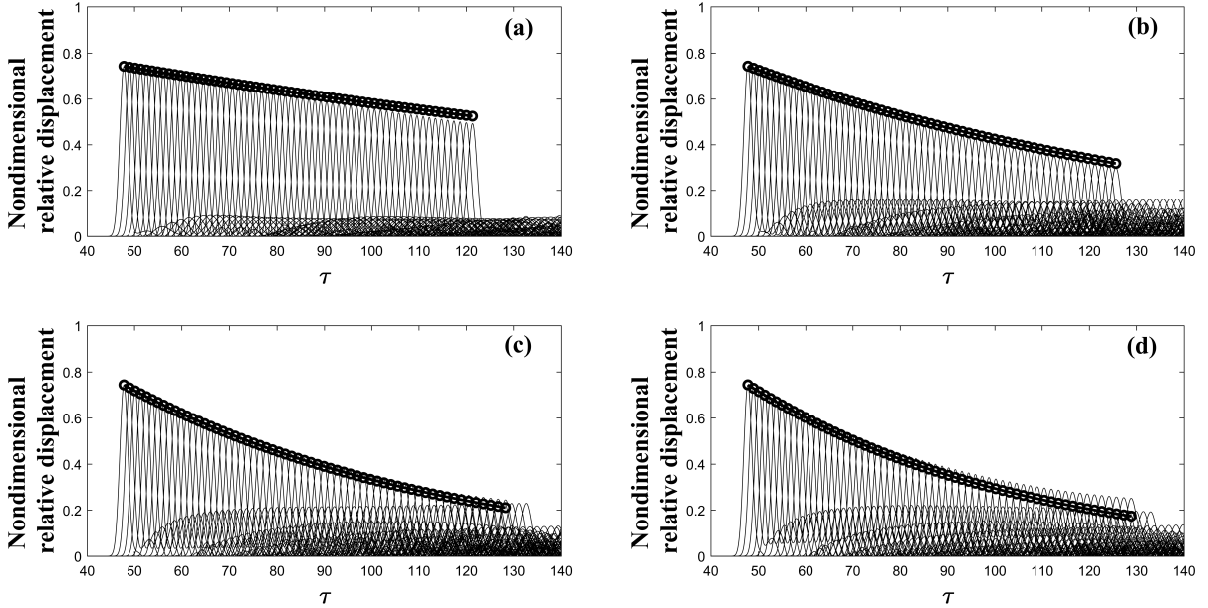}}
\caption{Time histories 
of the evolution of the relative displacements showing the
primary pulse in the 
second region, recorded on the first 60 contacts of the perturbed part of 
the chain. Results of numerical simulations are denoted with the solid 
lines; results of analytical approximation of (\ref{eq18}) are denoted with the connected 
circles (o). System parameters: $\nu =0.05$, (a) $\tilde \varepsilon =0.01$, (b) 
$\tilde \varepsilon =0.025$, (c) $\tilde \varepsilon =0.04$ and (d) $\tilde \varepsilon =0.05$.}
\label{refig8}
\end{figure}

\begin{figure}[htbp]
\centerline{\includegraphics[width=1.10\textwidth]{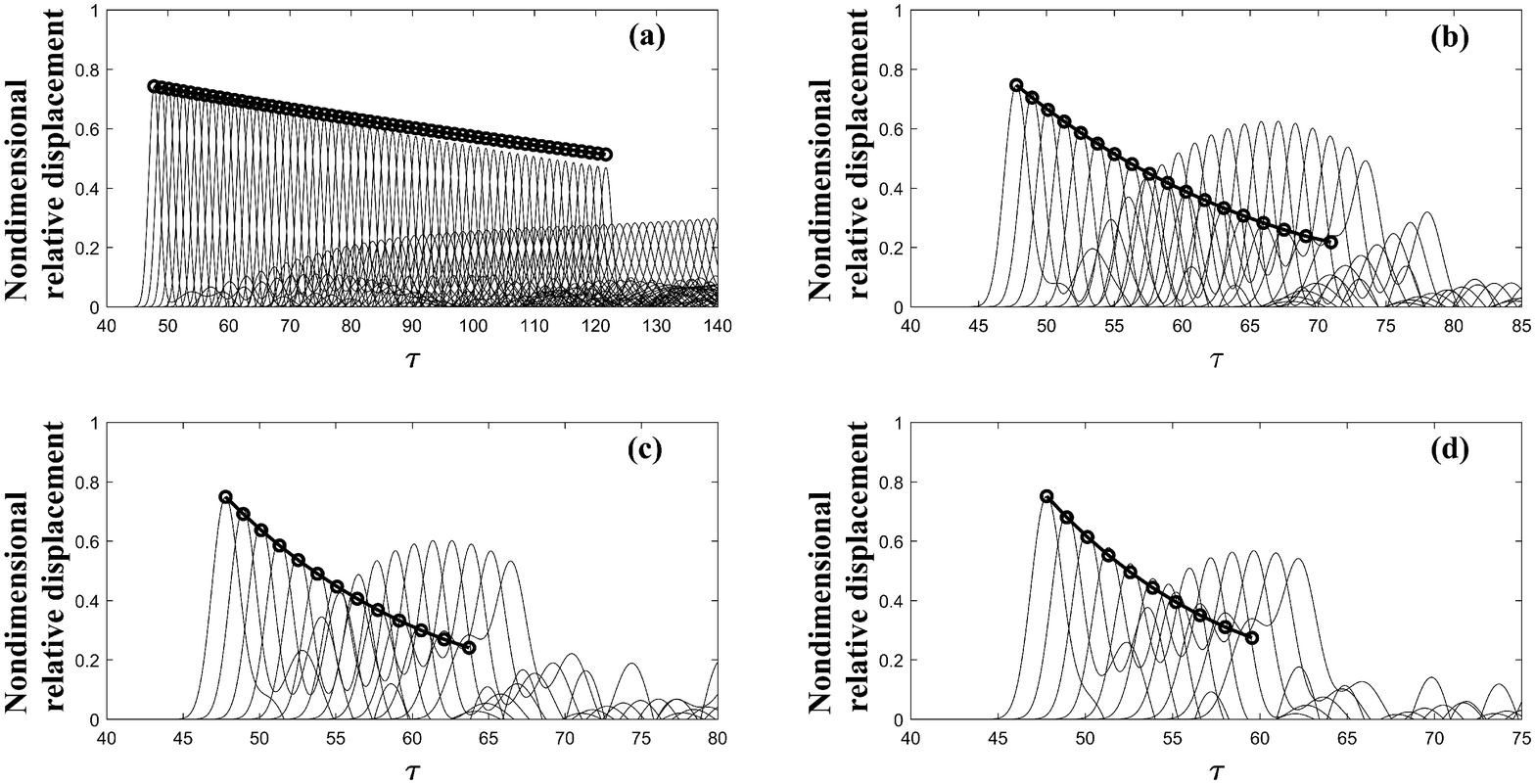}}
\caption{Same as Figure~\ref{refig8} for system 
parameters: $\nu =0.25$, (a) $\tilde \varepsilon =0.01$, (b) $\tilde \varepsilon =0.12$, (c) 
$\tilde \varepsilon =0.18$ and (d) $\tilde \varepsilon =0.25$.}
\label{refig9}
\end{figure}

Results of Figures~\ref{refig8} and \ref{refig9}
suggest that in the limit of small mass 
attachments and soft springs (i.e. $\tilde \varepsilon <1,\nu \sim O(\tilde \varepsilon 
))$ a strong rate of the decay of a single-humped primary pulse can be 
achieved. Moreover, as it is clear from the illustrated comparisons,
the analytical predictions for the decay of a primary pulse given by 
(\ref{eq18}) are 
found to be in very good correspondence with the results of the direct 
numerical simulations. It should be clear that in examining this
decay, we are disregarding the formation of the secondary humps,
which is an important feature in its own right that we will be concerned
with below.

A more careful look at the results shows that on one hand by changing the value of the internal stiffness $\tilde \varepsilon $ (staying in the limit of a soft internal spring and small mass attachment $0\le 
\tilde \varepsilon \le \nu \ll 1$), one can effectively control the rate of decay of a single-humped primary pulse. For example, as is shown in Figure~\ref{refig8} by fixing the value of $\nu=0.05$ and varying the stiffness parameter in the range ($0\le 
\tilde \varepsilon \le0.05$) one can clearly observe the increasing rate of decay of the primary pulse. However, on the other hand for the higher values 
of the internal stiffness, the distortion of the primary pulse resulting in 
the formation of the secondary hump occurs much faster. In other words the 
number of contacts of the second region sustaining the evolution of a 
single-humped primary pulse significantly decreases with the growth in the 
stiffness parameter. Thus, the further increase in the value of an internal 
stiffness parameter may lead to a more rapid formation of the localized, 
modulated wave-packets (multi-humped primary pulses) which also signals the
disappearance of the second region. Naturally, in that case 
the proposed analytical procedure cannot be  applied. 

As indicated above, when Region 2 exists and the pulse
decays sufficiently slowly, the analytical approximation provides
a highly accurate match to the decay of the principal pulse. 
Nevertheless, an additional question is whether the evolution
of the inner masses is also captured by the approximation,
as suggested e.g. by (\ref{eq108}). Unfortunately, the answer
to this question is in the negative, as clearly indicated in Figure~\ref{refig20}. Although, the approximation captures the
first stage of the passage of the primary pulse through the
bead (and hence its internal attachment), it subsequently 
considerably overestimates, through its ``rigid'' traveling pulse
assumption, the amplitude of the corresponding oscillation and
even misses its exact frequency. 
Notice, in addition, that the true system dynamics in the tails can 
become considerably more complicated e.g. including the separation of the outer 
spheres and emergence of chaotic motions.

\begin{figure}[htbp]
\centerline{\includegraphics[width=1.25\textwidth]{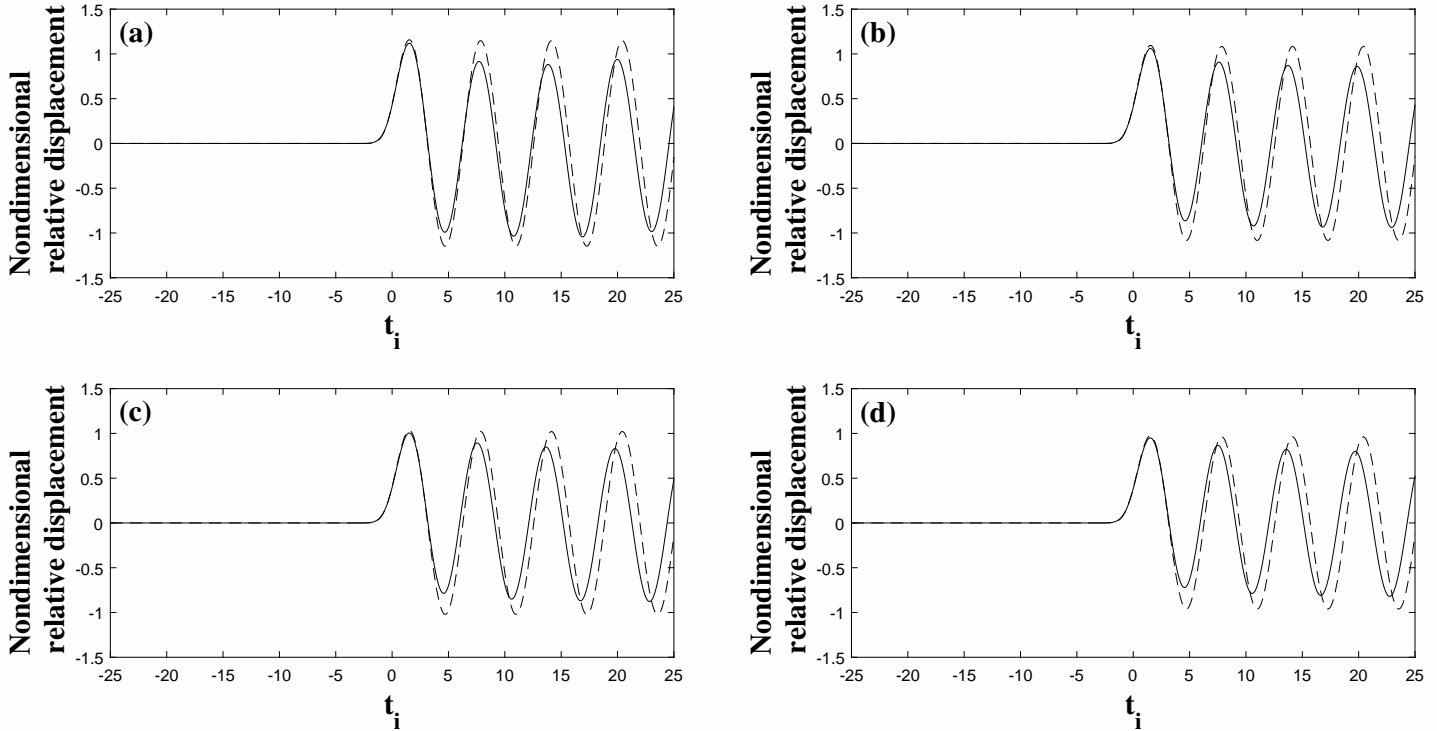}}
\caption{Comparison of the response of the relative displacements of 
the internal attachments ($d_{i}$) as derived from (\ref{eq108}) with the 
full direct 
numerical simulations. The 
response derived from (\ref{eq108}) is denoted with 
the dashed line; results of direct numerical simulations are denoted 
with the solid line. System parameters: $\nu =0.05,\mbox{\thinspace 
}\tilde \varepsilon =0.05$. (a) Response recorded on the 1st contact from the 
interface $d_{1} (t)$, (b) Response recorded on the 4th contact from the 
interface $d_{4} (t)$, (c) Response recorded on the 7th contact from the 
interface $d_{7} (t)$ and (d) Response recorded on the 10th contact from the 
interface $d_{10} (t)$.}
\label{refig20}
\end{figure}

{\bf Case (ii):}

From the discussion of Section III it is quite clear that for the cases of 
$\nu =O(1)$ or $\nu \gg 1$, for hard excitations the primary response of 
the system can be effectively approximated by a 
simplified model, namely by the uniform granular chain mounted on a linear 
elastic foundation. This type of the system has been studied analytically 
and numerically in~\cite{ref26}. 
In the same study it was shown that increasing the 
strength of elastic foundation (i.e. the value of the coefficient of 
stiffness) we increase the rate of the decay of the primary pulse. In the 
present study we demonstrate that the decay of the primary pulse in the case 
of the heavy mass attachments can be accurately approximated by the
analysis of Section III. 

To this end, we plot the drop in the amplitude of the primary pulse recorded 
on the first 10 contacts in Region 2 of the perturbed granular chain for the fixed value 
of the stiffness coefficient ($\tilde \varepsilon =0.1)$ and various values of the 
internal masses $(\nu =1, 3, 5, 10)$ in Figure~\ref{refig21}. 
The results suggest a very good 
correspondence of the analytical prediction of the drop 
 with that of the numerical simulation.

\begin{figure}[htbp]
\centerline{\includegraphics[width=1.20\textwidth]{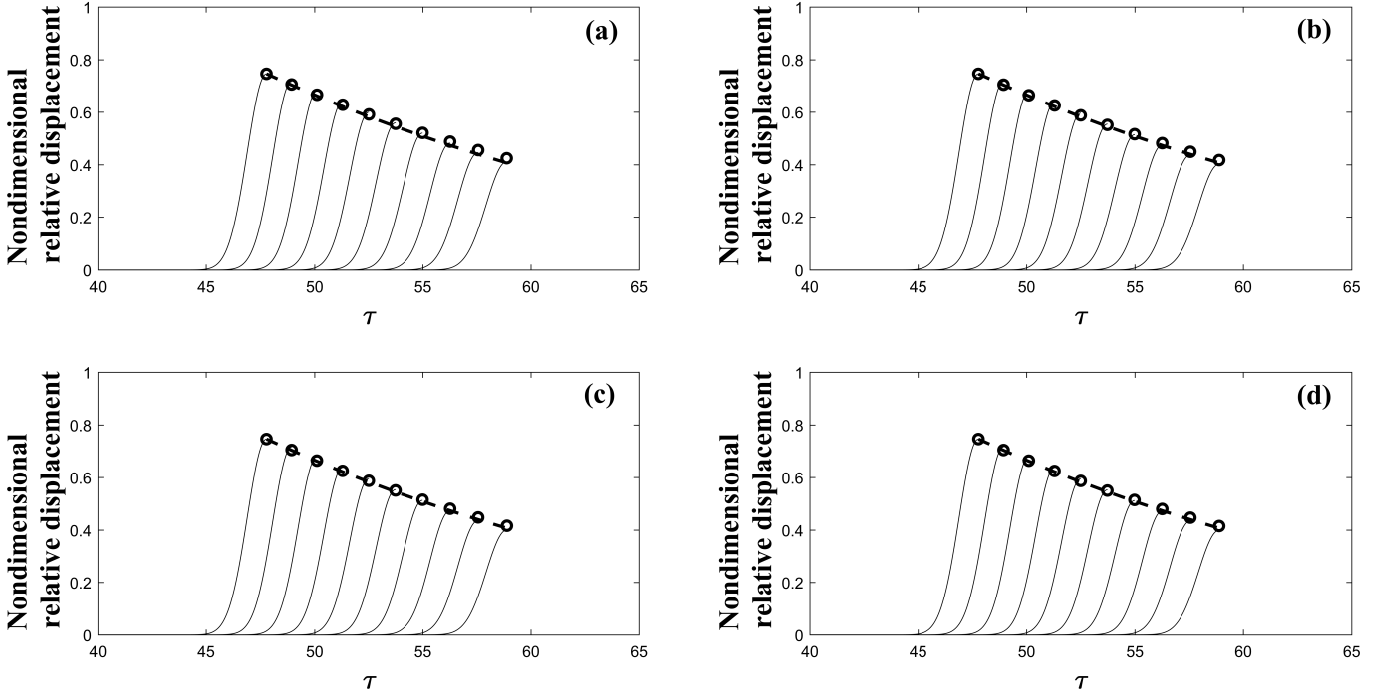}}
\caption{Comparison of the analytical prediction and numerical 
simulation of the primary pulse amplitude decay on the outer spheres for 
$\tilde \varepsilon =0.1$ (a) $\nu = 1$ (b) $\nu = 3$ (c) $\nu = 5$ (d) $\nu 
= 10$ in the non-dimensional relative displacement recorded on the contacts.
The symbols denote the amplitudes of the 
primary pulse in terms of the relative displacement predicted by the map 
(\ref{eq18}), while the 
dashed line shows the corresponding 
long wave approximation of (\ref{eq22}). The bold black lines correspond to full
lattice numerical simulations.} 
\label{refig21}
\end{figure}

{\bf Case (iii): }

In Figure~\ref{refig22}, 
we illustrate the formation of moving breathers for the low 
amplitude excitations and the case of heavy linear attachments.
This feature serves to showcase the accordance of the full numerical
simulation to  
the reduced order model (\ref{eq26}), which supports such structures,
as analyzed in the works of~\cite{ref28,ref29}. Hence, we do not
pursue this avenue further, but defer the interested reader to
the detailed structures in such equations as presented
in the above works.

\begin{figure}[htbp]
\centerline{\includegraphics[width=1.00\textwidth]{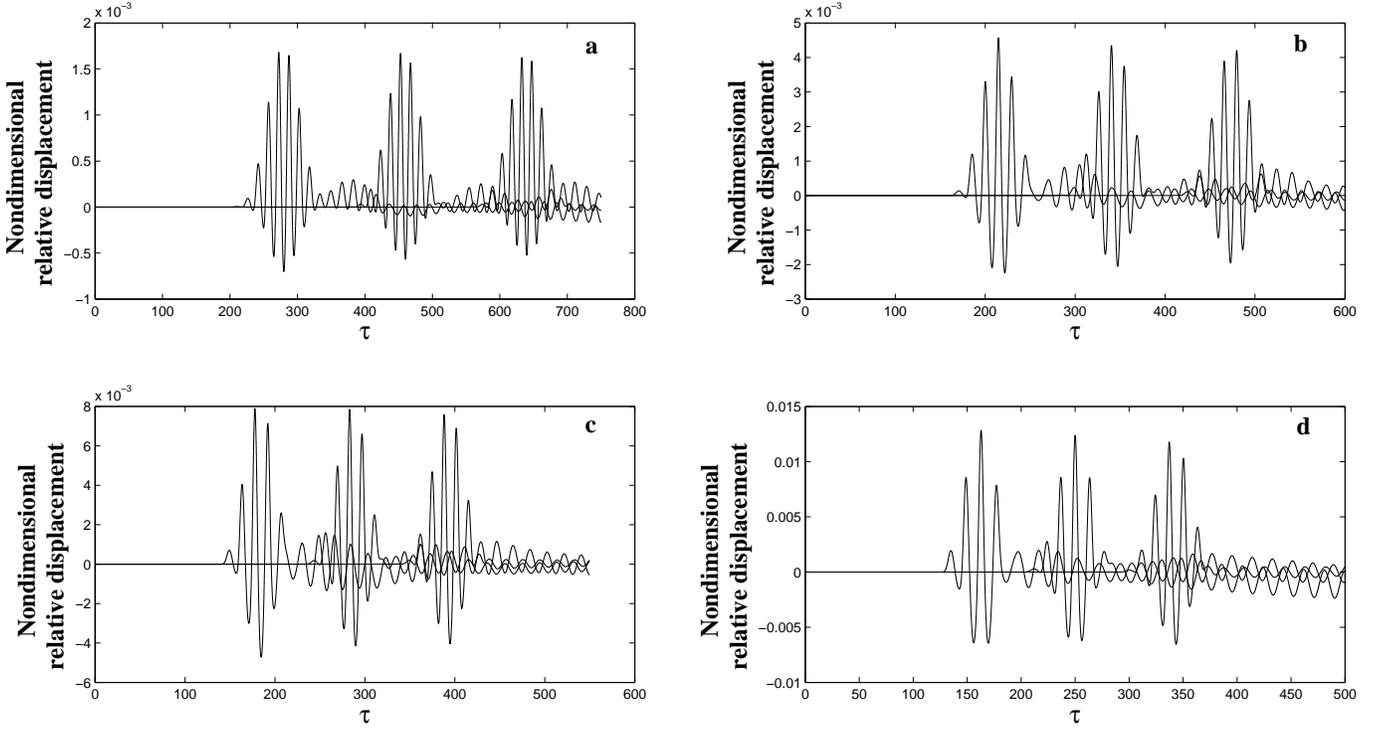}}
\caption{The non-dimensional relative displacement recorded on the 
contacts from the direct numerical simulation of (\ref{eq6})$_{\mathrm{\thinspace 
}}\varepsilon =0.1$ (a) $\nu =1$ (b) $\nu =3$ (c) $\nu =5$ (d) $\nu =10$.}

\label{refig22}
\end{figure}

\subsection{Efficiency of Attenuation and the Formation of 
Robust Traveling Waves}

To study the efficiency of the local resonators in inducing attenuation of the 
primary (single humped) pulses as they propagate through the perturbed part 
of the chain, one needs first to define a precise criterion. 

Here, we have chosen to focus on the maximal velocity of the last element after its detaching from the chain. In each numerical run we set the initial velocity of the first element to unity (i.e. $V_{0} = 1$) while the total number of the elements in the chain is fixed to $N = 125$ (here we note again that the first 39 elements contain no internal resonators while the rest elements of the chain incorporate, perfectly identical local resonators). In each numerical run we evaluate the maximal velocity of the last element $V_{N}\left( {\tilde \varepsilon ,\nu } \right)$ after its complete detachment from the chain. It is worthwhile noting that the calculated data (i.e $V_{N}\left( {\tilde \varepsilon ,\nu } \right)$ for $0 < \tilde \varepsilon , \nu \le 1$) is normalized with respect to the maximal velocity of the latter in the given range of the system parameters ($0 < \tilde \varepsilon , \nu \le 1$) (See Figure~\ref{refig5}). 

\begin{figure}[htbp]
\centerline{\includegraphics[width=10cm]{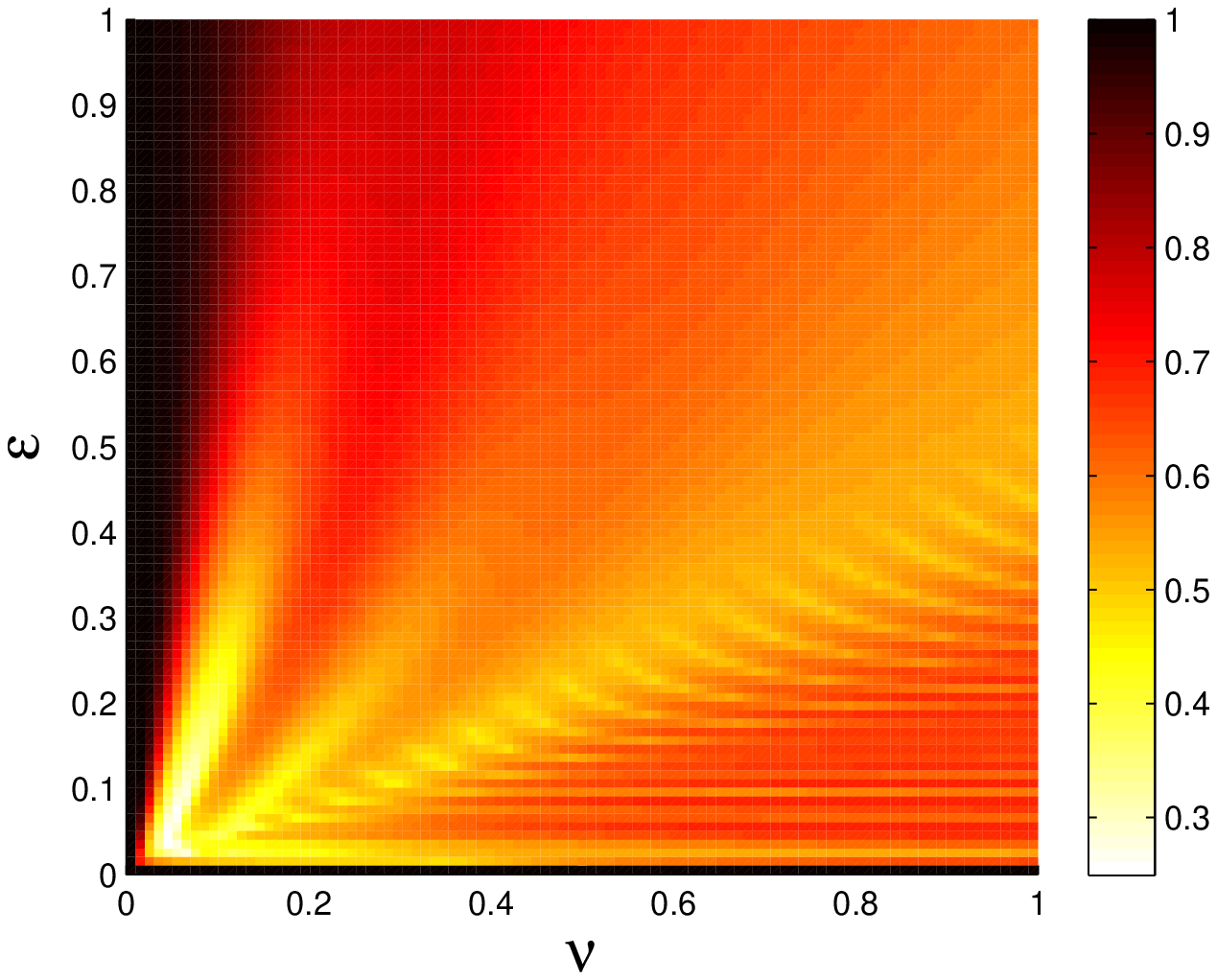}}
\caption{Contour map of the attenuation diagnostic $V_{N}\left( {\tilde \varepsilon ,\nu } \right)$ in the 
range of $0<\tilde \varepsilon ,\nu <1$. The results correspond
to full lattice numerical simulations.}
\label{refig5}
\end{figure}

From the observation of the results of Figure~\ref{refig5},
it is clear that there is a 
certain tradeoff between the mass and the stiffness of the internal 
attachment leading to the best possible attenuation of the primary pulse. 
Moreover, a close look at the results of Figure~\ref{refig5} 
shows that the area of  
efficient attenuation is achieved for the relatively small values of the 
internal mass and stiffness coefficient. In particular, the very efficient attenuation is reached for $\tilde \varepsilon \simeq 
0.05,\nu \simeq 0.05$. In addition, one can observe the formation of the new regions corresponding to the case of the relatively heavy internal masses e.g. $\left(\tilde \varepsilon \simeq 
0.05,\nu \simeq 0.9\right){}$ leading to the efficient attenuation of the primary pulse. 

Another very  
interesting observation made during the numerical study of the 
regimes corresponding to the third region is the formation of what appears
to be 
stationary solitary like pulses of various shapes. In Figure~\ref{refig6},
 we plot an example of a  
single humped such pulse (left panel), as well as one of 
a double humped solitary wave (right panel), respectively which have been 
formed in the third region. Notice that in the latter, it is the hollow
spheres that feature the double-humped response in time, while the inner
attachment is still single-humped. 

\begin{figure}[htbp]
\centerline{%
\includegraphics[width=0.575\textwidth]{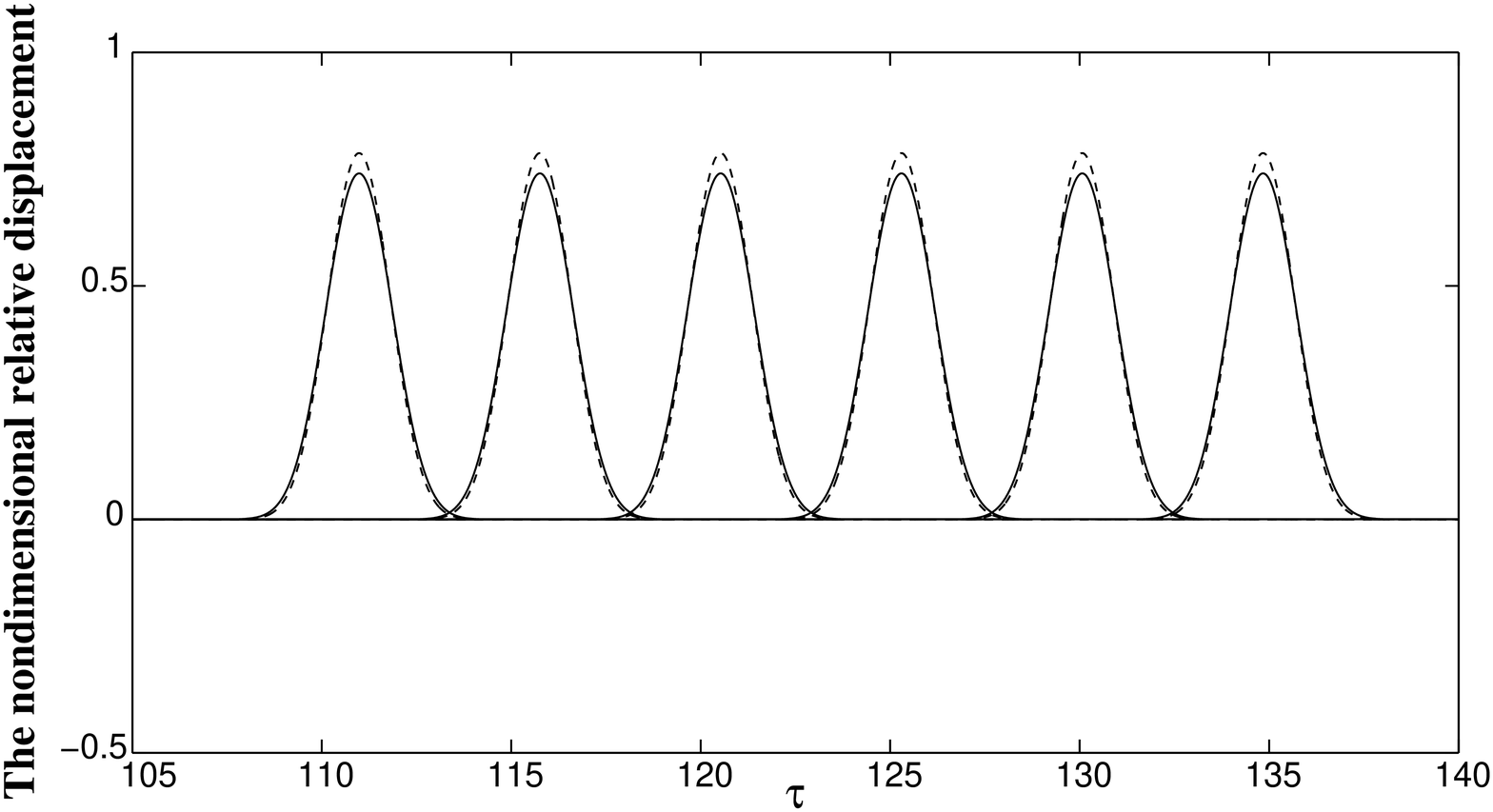}
\includegraphics[width=0.575\textwidth]{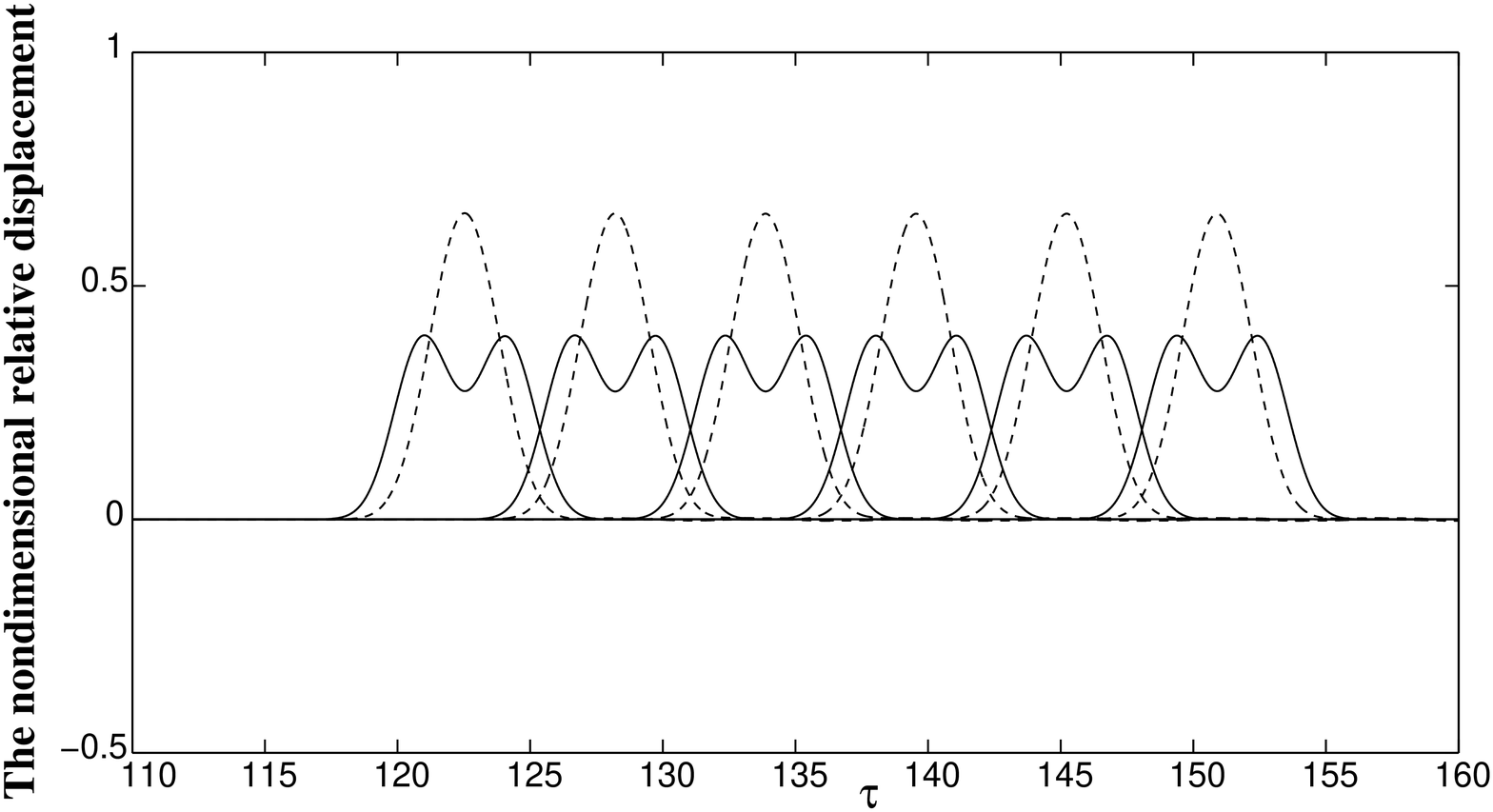}}
\caption{Time histories from numerical simulations 
illustrating a single-humped solitary like pulse 
recorded on the six different contacts (92,94,96,98,100,102) in the third region
(left panel). 
 The 
system parameters are $\tilde \epsilon 
=0.105$ and $\nu =0.004$.
A similar example but now for a double-humped solitary like pulse 
recorded on the six different contacts (92,94,96,98,100,102) in the third region
is shown in the right panel. Here, the system parameters are $\tilde \epsilon 
=0.105$ and $\nu =0.10245$. In both cases, the 
solid line denotes the response of the outer masses, while
the  dashed line denotes 
the response of the inner masses.}
\label{refig6}
\end{figure}

These observations have led us to a more systematic study of such a
response which is reflected in Figures~\ref{fig16a} and \ref{fig16b}. 
The former is for the case $\tilde \varepsilon 
=0.105,\nu =0.10245$, while the latter for $\tilde \epsilon 
=0.105,\nu =0.004$. In these more detailed, larger scale runs,
the wave is initialized at bead $1$ 
while the interface is at
bead $40$ and all beads carry a local resonator thereafter.
The space time plot in the top panel of Figure~\ref{fig16a}
(as well as in the left panel of Figure~\ref{fig16b})
already spells out the fundamental phenomenology. The initial excitation
induces the formation of the traveling wave (notice that additional
excitations stem from the reflection of this bead at a later stage).
At the interface, a significant amount of excitations is produced which
is visible (and is sustained thereafter) in the space-time contour
plot of the velocity field $\tilde X_{i,\tau}$. This leads to the decay
of the primary pulse, and the formation of a secondary one. 
However, shortly thereafter, the interference of the primary and
secondary pulse distill a coherent waveform that appears to detach
itself from residual ``radiation'' and to travel faster than the latter
towards beads of higher index $i$. Observation of the contour plots,
as well as of Figure~\ref{refig6} initially appears to suggest that the
solitary wave is a genuinely localized one.

Nevertheless, this observation is {\it misleading}, as is revealed
by the bottom panels of Figure~\ref{fig16a} and the right panel
of Figure~\ref{fig16b}, In particular, the same features as indicated
above in Figure~\ref{refig6} are evident in the evolution of the time
derivative of the displacement at the sites $i=200$, $i=210$ and $i=220$.
Yet, there are also clearly discernible tails which are highly ordered
and follow the primary pulse. As the arbitrary oscillations, shown
for comparison by the dash-dotted lines, suggest, the tail of the
wave oscillates with {\it precisely} the natural linear frequency
of the system, namely the frequency of the out-of-phase oscillator
of the outer shell and its inner linear attachment, namely
with $\omega=\sqrt{\kappa (1+1/\nu)}$. This is true for both
examples (with very distinct linear frequencies due to the 
difference in the values of $\nu$) of the bottom panels of
Figure~\ref{fig16a} and the right panel
of Figure~\ref{fig16b}.

\begin{figure}[tbp]
\centerline{
\includegraphics[width=11.cm]{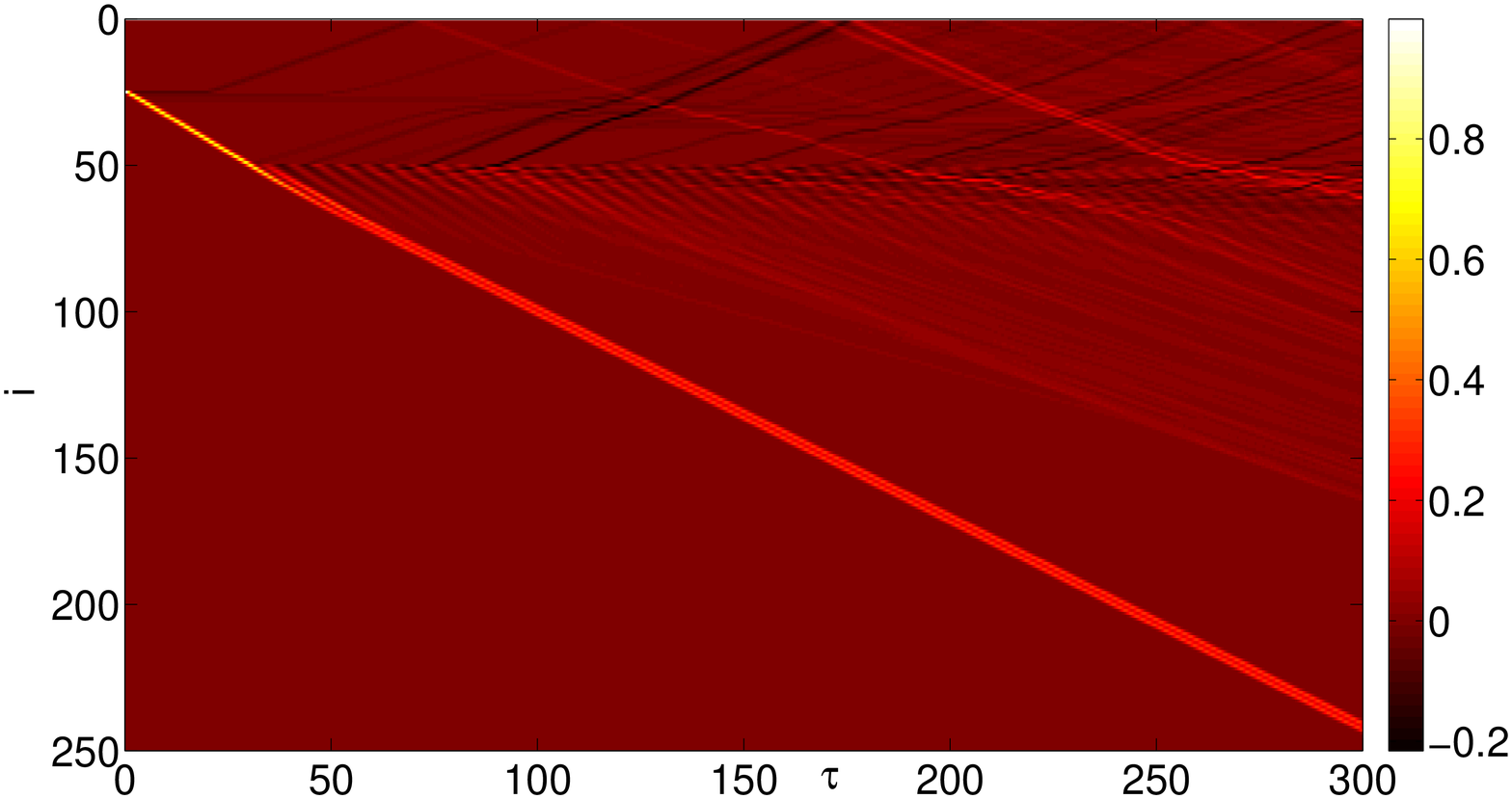}
}
\centerline{
\includegraphics[width=9.cm]{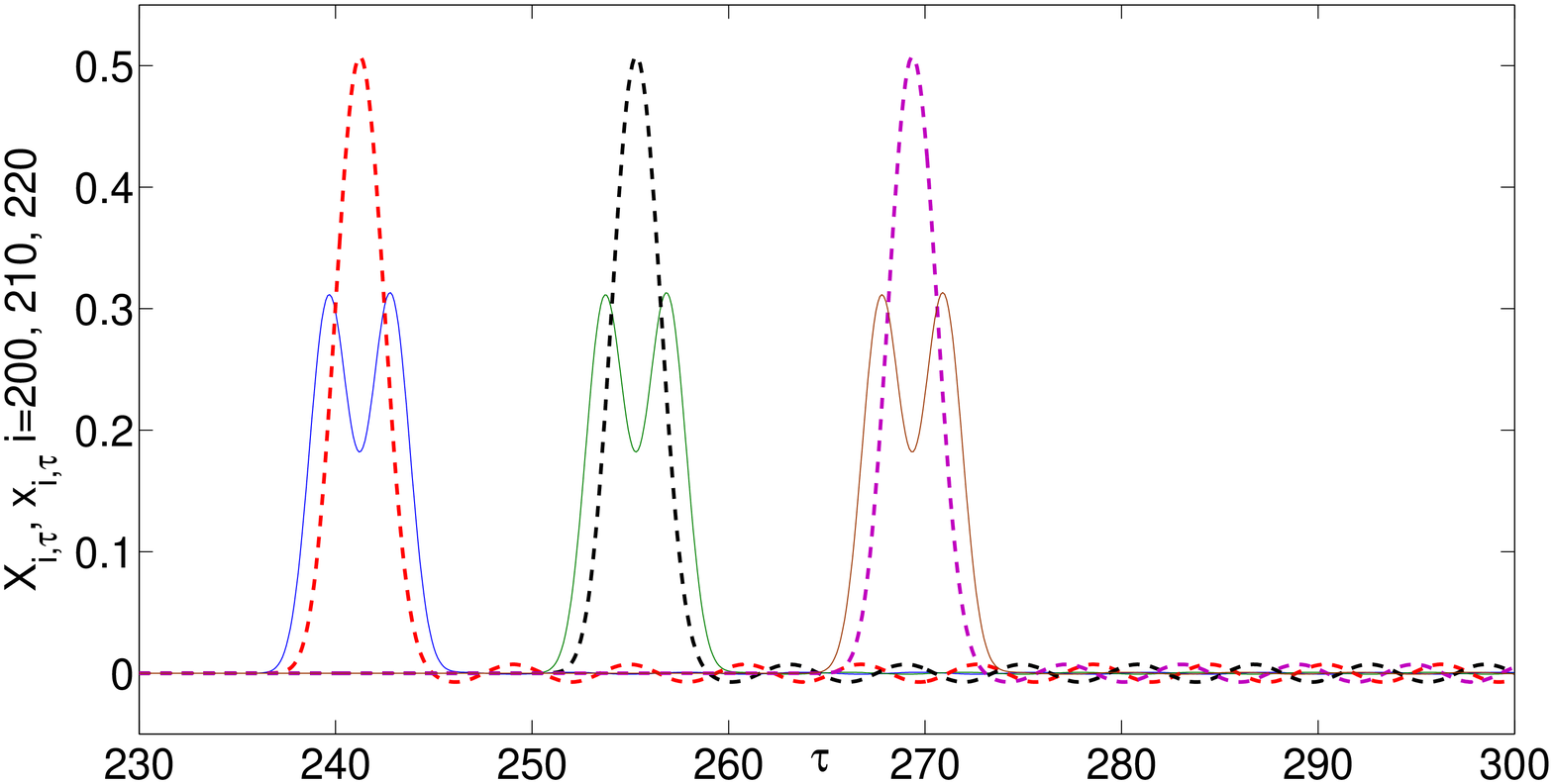}
\includegraphics[width=9.cm]{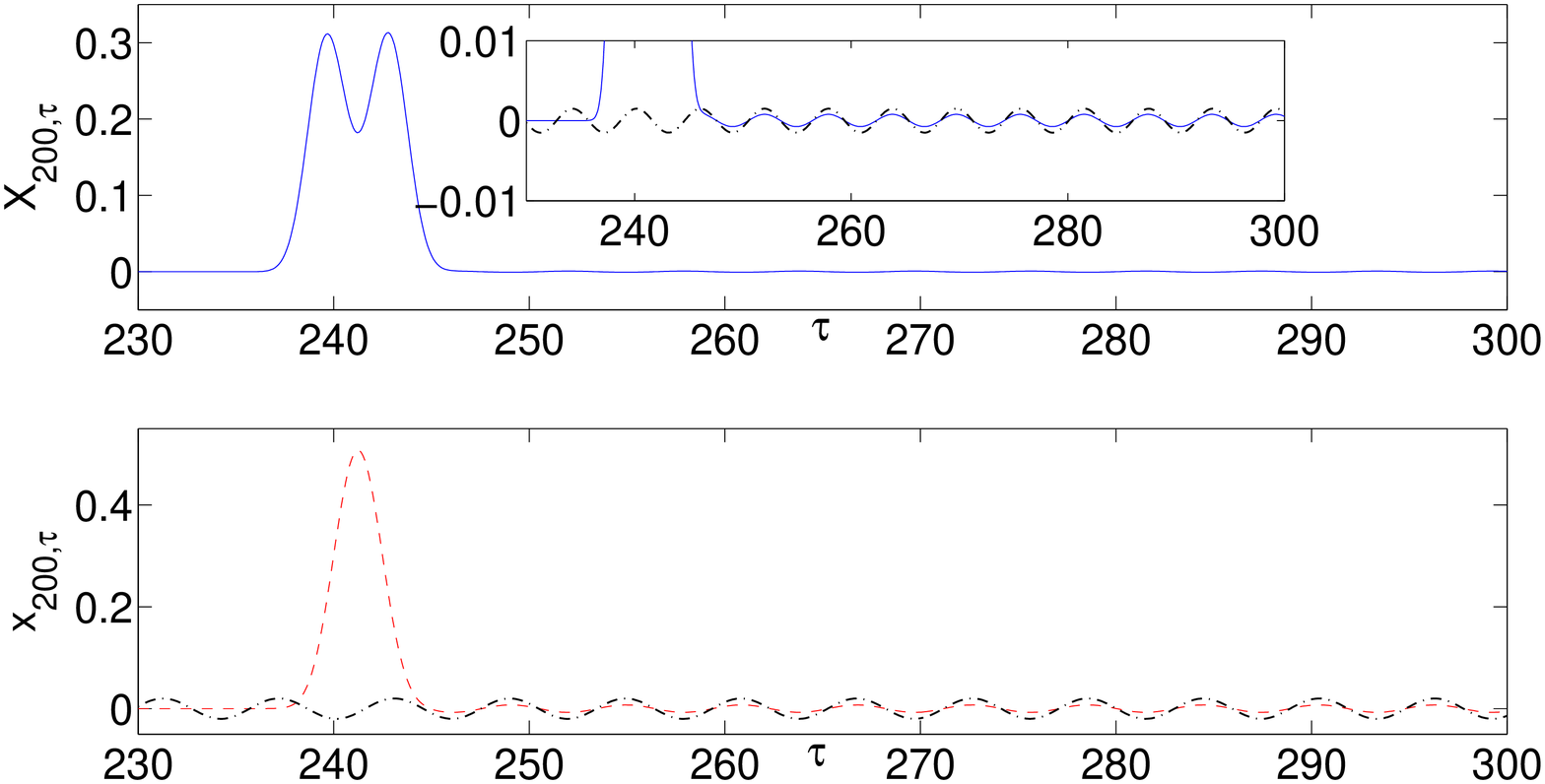}
}
\caption{The top panel shows the space-time contour plot
of $\tilde X_{i,\tau}$, the time derivative of the displacement of the
outer shell of our granular chain for $\tilde \varepsilon 
=0.105$ and $\nu =0.10245$. The particular evolution for 3
sites ($i=200$ leftmost, $i=210$ middle and $i=220$ rightmost)
are shown in the bottom left panel. The velocity of the the outer shells are given
by the solid lines, while the dashed ones provide the evolution
of the inner attachments. The presence of the oscillating tails
is evident and is rendered more transparent in the right panel
for site $i=200$, where the outer shell field $\tilde X_{i,\tau}$ (top)
is separated from the inner attachment field $\tilde x_{i,\tau}$ (bottom).
Both of them oscillate and for comparison a dash-dotted sinusoidal
curve with amplitude comparable to their tail oscillation amplitude and
frequency selected as $\omega=\sqrt{\kappa (1+1/\nu)}$ is explicitly given
to illustrate indeed the precise agreement of the tail oscillation 
with the frequency of the out-of-phase motion between the inner
and outer beads. For the outer shell, the oscillation is not visible
on the scale of the plot, and hence is given in the relevant inset
which is a blowup at a proper scale to render it visible.}
\label{fig16a}
\end{figure}

\begin{figure}[tbp]
\centerline{
\includegraphics[width=9.cm]{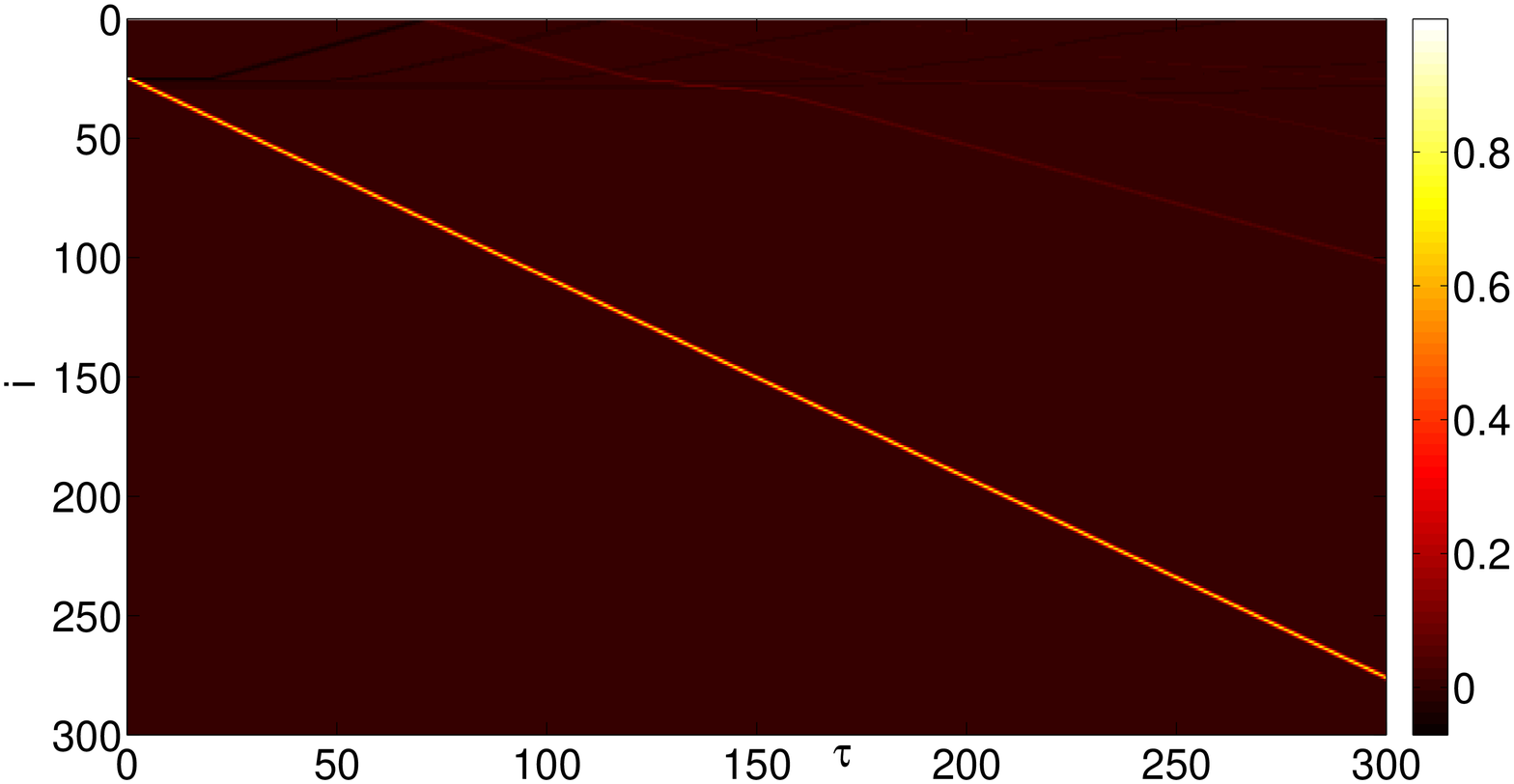}
\includegraphics[width=9.cm]{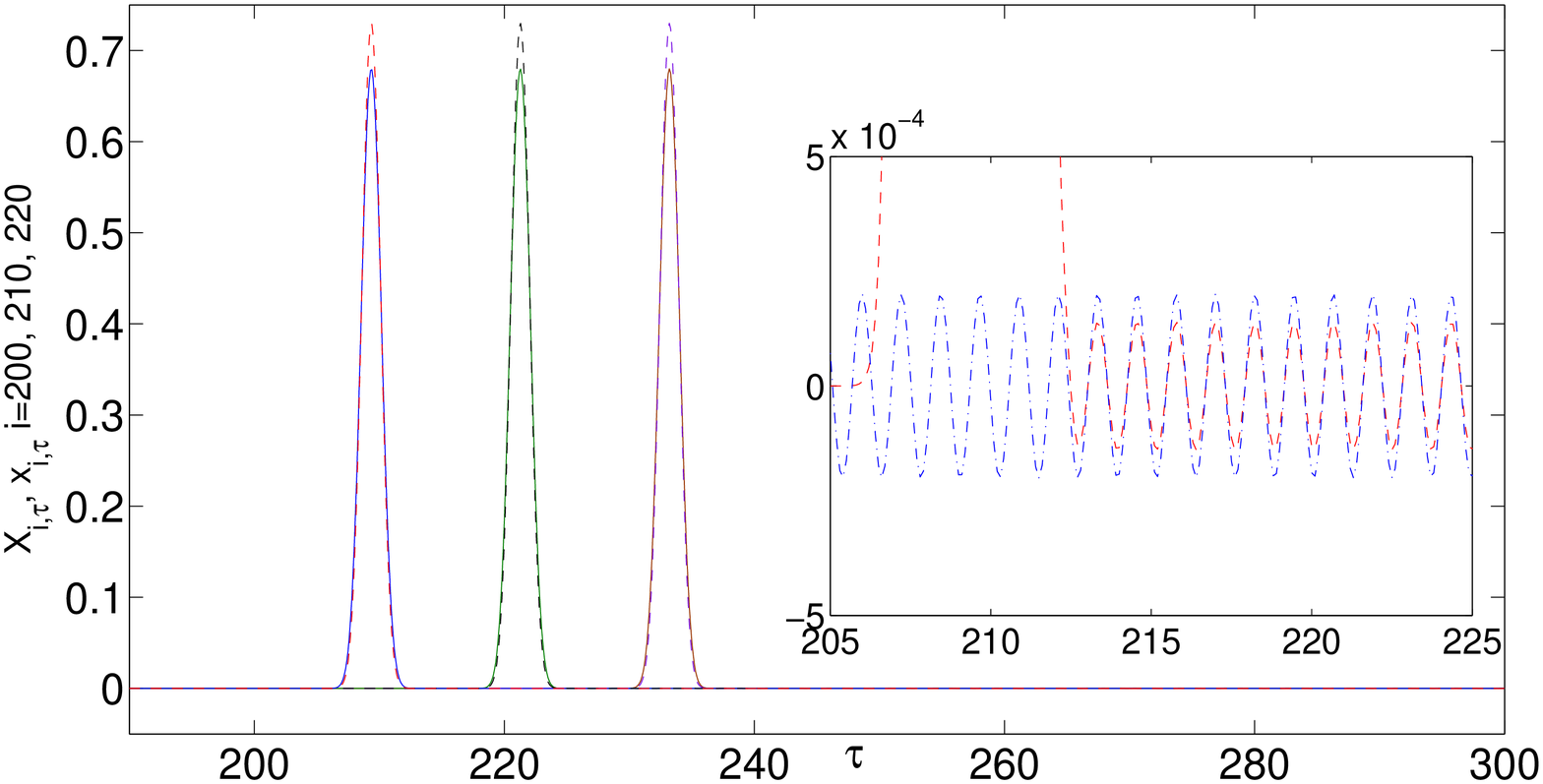}
}
\caption{Same as the previous figure, but now for the case
of the single humped primary pulse for parameters 
 $\tilde \epsilon 
=0.105,\nu =0.004$. In this case, as well, the right panel
and specifically its inset and its comparison to an arbitrary
sinusoidal curve with frequency  $\omega=\sqrt{\kappa (1+1/\nu)}$
allow us to appreciate the existence of a tail oscilating
with the above linear frequency.}
\label{fig16b}
\end{figure}

The above features constitute a significant new development.
It is clear from the figures that an excitation propagates
in the chain {\it without visible distortion} or attenuation, 
being preserved as a {\it weakly nonlocal solitary wave}.
The latter waves, also known as nanoptera, constitute a type of
waveform that in addition to a strongly localized
core bears an extended tail. This is a theme of research
that has attracted a considerable volume of activity 
from a wide range of areas including fluid
mechanics, nonlinear optics and oceanography among
others~\cite{boyd,review}, as well as from the mathematical
physics associated with exponential (and other types of) 
asymptotics~\cite{review,montaldi}. 

Hence, we devote
the next section to an attempt to produce such coherent structures
as exact traveling waves in our mass-in-mass chains.  

\section{Nanopteronic Solutions in the System}

Here, we will showcase a proof-of-principle example of the analysis
and computation of nanopteronic waves of the type identified
in the previous section, as well as of its uninhibited
propagation through the granular crystal. In what follows, we will
focus on nanoptera which have symmetric tails on both the ``front''
and the ``back'' of the main core, but the
numerical computations of the previous section suggest
that such entities can be quite long-lived
even in the asymmetric case where they bear a ``single arm''.
More specifically, the nature of our
initial conditions in section IV is not conducive
to the formation of weakly localized nonlinear waves with such
symmetric tails, but only to ones with a single arm/tail. 
However, as will be evident also from the
numerical results herein, this is inconsequential as the core
and the tail travel at the same speed, forming a coherent traveling
structure.

Let us focus on the system of equations for the strains (where
for simplicity we have set $\kappa=1$, although this can straightforwardly be restored):
\begin{eqnarray}
{\Delta}_{i,\tau\tau} &=&  
\left(\Delta_{i+1,+}^{3/2} + \Delta_{i-1,+}^{3/2} -2 \Delta_{i,+}^{3/2} \right) -  
(\Delta_i - d_i)
\label{eqn3}
\\
\nu {\delta}_{i,\tau\tau} &=& - \left( d_i - \Delta_i \right).
\label{eqn4}
\end{eqnarray}

If we consider waves that spatially decay, in the regime of 
small amplitude, we can always identify wave amplitudes $\Delta_i$
(and $\delta_i$) small enough such that $\Delta_i \gg \Delta_{i}^{3/2}$. Hence,
it will always be the case that for small enough amplitudes, the
relevant equations governing the evolution of the wave can be
approximated as ${\Delta}_{i,\tau \tau}=-(\Delta_i-d_i)$ and 
${d}_{i,\tau \tau}=-(1/\nu) (d_i-\Delta_i)$.
These equations lead to $d_i=-\Delta_i/\nu$ and
$\Delta_i$ satisfies the effective oscillator equation 
${\Delta}_{i,\tau\tau}=-(1+1/\nu) \Delta_i$. Hence, it is natural to expect
for small amplitude vibrations left behind a propagating 
wave with an oscillation of temporal frequency
$\omega=\sqrt{1+1/\nu}$; in the presence of $\kappa$,
this frequency becomes $\sqrt{\kappa (1+1/\nu)}$.

\subsection{The Pego-English Approach and the Existence of Nanoptera}

We now seek traveling waves of speed $c$ to the dynamical equations in the
form $\Delta_i(t)=R(i-c t)$ and $d_i(t)=S(i-c t)$,
substituting these expressions in (\ref{eqn3}) and (\ref{eqn4}).
The resulting 
advance-delay differential equations in the traveling wave
variable $\xi=i-c t$ are then of the form:
\begin{eqnarray}
c^2 R''(\xi) &=& \left[ R(\xi+1)_+^{3/2} + R(\xi-1)_+^{3/2} -2 R(\xi)_+^{3/2} \right]
- \left( R(\xi) - S(\xi) \right)
\label{eqn5}
\\
c^2 \nu S''(\xi) &=& - \left( S(\xi) - R(\xi) \right),
\label{eqn6}
\end{eqnarray}
where the prime denotes derivative with respect to its argument.
We now use the Fourier transform of these equations according to the
definition
$\hat{R}(k)=(1/2 \pi) \int_{-\infty}^{\infty} R(\xi) e^{i k \xi} d \xi$
(and similarly for $S$).
We thus obtain in this generalization of the approach used in~\cite{english} 
the following conditions (upon writing the relevant algebraic
equations and solving them for $\hat{R}$ and $\hat{S}$) 
\begin{eqnarray}
\hat{R} &=& \frac{1-\nu c^2 k^2}{1+\nu - \nu c^2 k^2}
\left(\frac{\sin(\frac{k}{2})}{c \frac{k}{2}}\right)^2 \widehat{(R^{3/2})}
\label{eqn7}
\\
\hat{S} &=& \frac{1}{1- \nu c^2  k^2} \hat{R}.
\label{eqn8}
\end{eqnarray}
As a side note, we should mention here that in the infinite lattice 
the above equations should be considered in the sense of distributions,
given the non-vanishing nature of the tails. Nevertheless, in what
follows, we will restrict our considerations to a finite domain.

The above equations can be considered as an iterative scheme
that can be used computationally in order to retrieve its
fixed point (nontrivial) solution that would constitute the traveling
wave of the original dynamical (strain) equations. In particular, 
given a profile $R$, the right hand side of (\ref{eqn7}) 
is computed and used to obtain the next order guess for
$\hat{R}$, whose subsequent use in (\ref{eqn8}) and
Fourier transform of both $\hat{R}$ and $\hat{S}$ will
give rise to the next order guesses, and so on, till convergence.
Naturally, for $\nu=0$, this iteration scheme
retrieves the one used to obtain the solitary waves
of the monoatomic granular chain in~\cite{english}.

An additional possibility which would avoid the complications
(slow tail decay in Fourier space) of the Fourier calculation
proposed above would be to revert (\ref{eqn7}) and (\ref{eqn8})
into real space. That would reduce the number of modes needed
for an accurate computation of the solution.
In the present computation, a large number of modes is used
over domains of size $120$ or even $240$, with $dx=0.01$
and periodic boundary conditions (although only a small
fraction of the lattice is shown for clarity).

A typical example of the result of the iteration scheme
of (\ref{eqn7}) and (\ref{eqn8}) is given in Figures~\ref{fig1} and \ref{fig2}
for the case of $\nu=0.05$. Figure~\ref{fig1} illustrates the
result of the iterative scheme of the equations towards the 
acquisition of a traveling wave solution.

It is not only interesting that
the scheme continues to converge for this finite value of 
$\nu$. It is, in fact, remarkable that, as is partially anticipated
from the argument of the previous sections (and from our results
illustrating the attenuation of a regular monotonic pulse and the
spontaneous emergence of such weakly nonlocal solitary waves),
the obtained exact solutions have a {\it nanopteronic} character.
Namely, they feature spatial oscillations with a period
which is consonant with the dominant wavenumber (which is
$(1/c) \sqrt{1+1/\nu}$ as illustrated in (\ref{eqn7})). 
In the latter equation, the poles along the real $k$-axis
arise precisely at $k_0=(1/c) \sqrt{1+1/\nu}$, suggesting that
the corresponding Cauchy principal value integral (arising when Fourier
transforming to obtain $R(\xi)$), will naturally give rise
to a sinusoidal dependence on $k_0 \xi$. From (\ref{eqn8}),
this is expected to be reflected in the spatial dependence of
$S(\xi)$ as well. This, in turn, is consonant with the temporal
oscillation featuring the frequency of the out-of-phase mode,
as the corresponding frequency $\omega_0=c k_0$ and hence is in 
agreement with the above numerical observations of 
Figures~\ref{fig16a} and \ref{fig16b}.
This implies that there exists a resonance between the
propagation of the traveling wave and the excitation
of the linear out-of-phase mode of each resonator,
a feature that has been analyzed in detail in~\cite{haitao}.
In fact, in the latter work, additionally a so-called
antiresonance condition (of $k=2 n \pi$) was also identified
under which a regular solution with monotonically decaying
tail was identified. 
Only in this case of anti-resonance has it recently been
proved that a solution in this resonator chain exists
(and is bell-shaped)~\cite{haitao2}. Admittedly, the proof
of convergence of the above scheme in the
case of nanoptera with non-vanishing tails is an important
open problem (as is the rigorous mathematical proof of existence
of such states). 

While
$R(\xi)$  and $S(\xi)$ are obtained as continuum solutions in the
left panel of the figure, they are ``distilled'' on the lattice
in the right panel and so is the corresponding lattice momentum
associated with the chosen speed of $c=1$. 

\begin{figure}[tbph]

\centerline{
\includegraphics[width=9.cm]{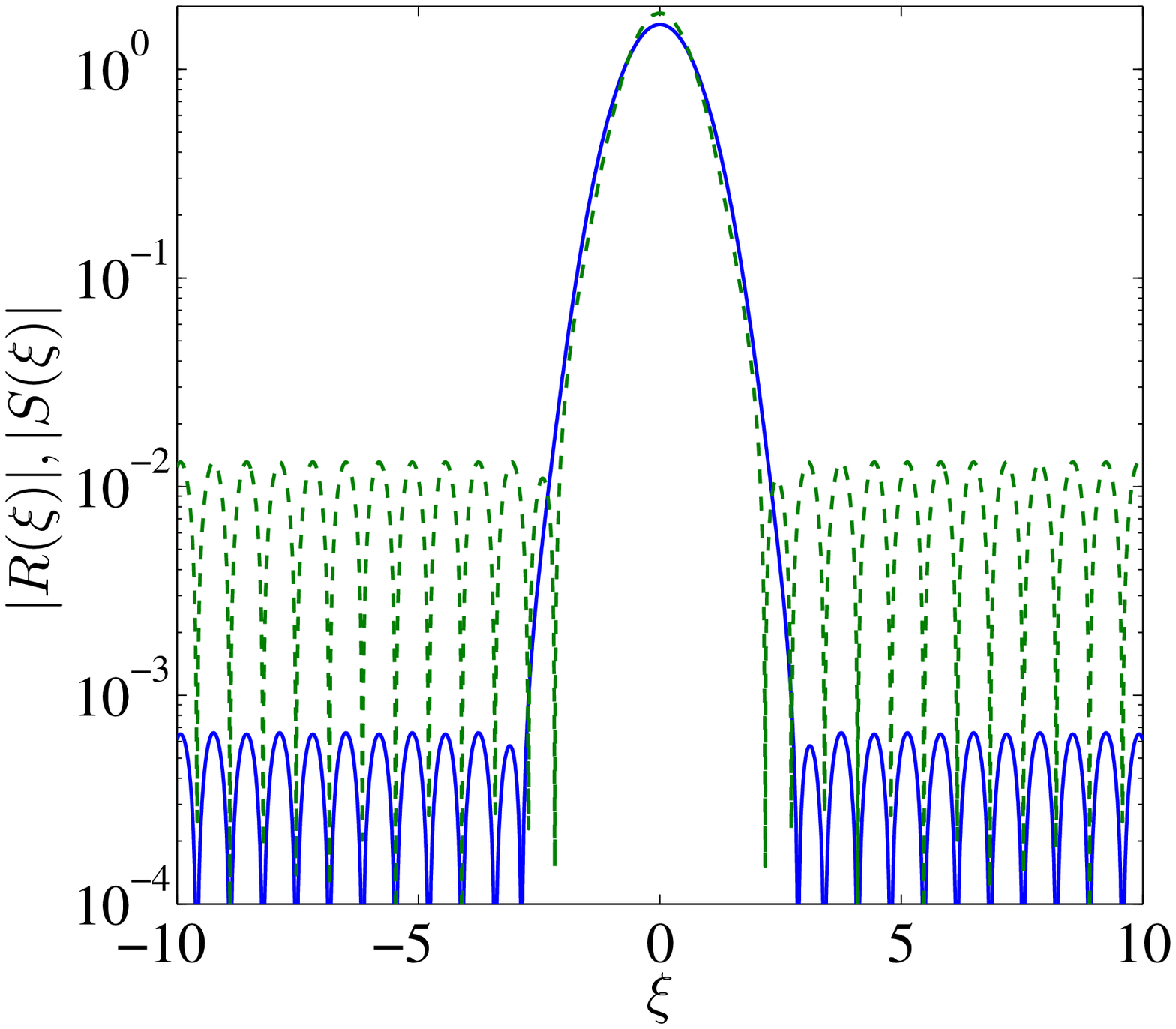}
\includegraphics[width=9.cm]{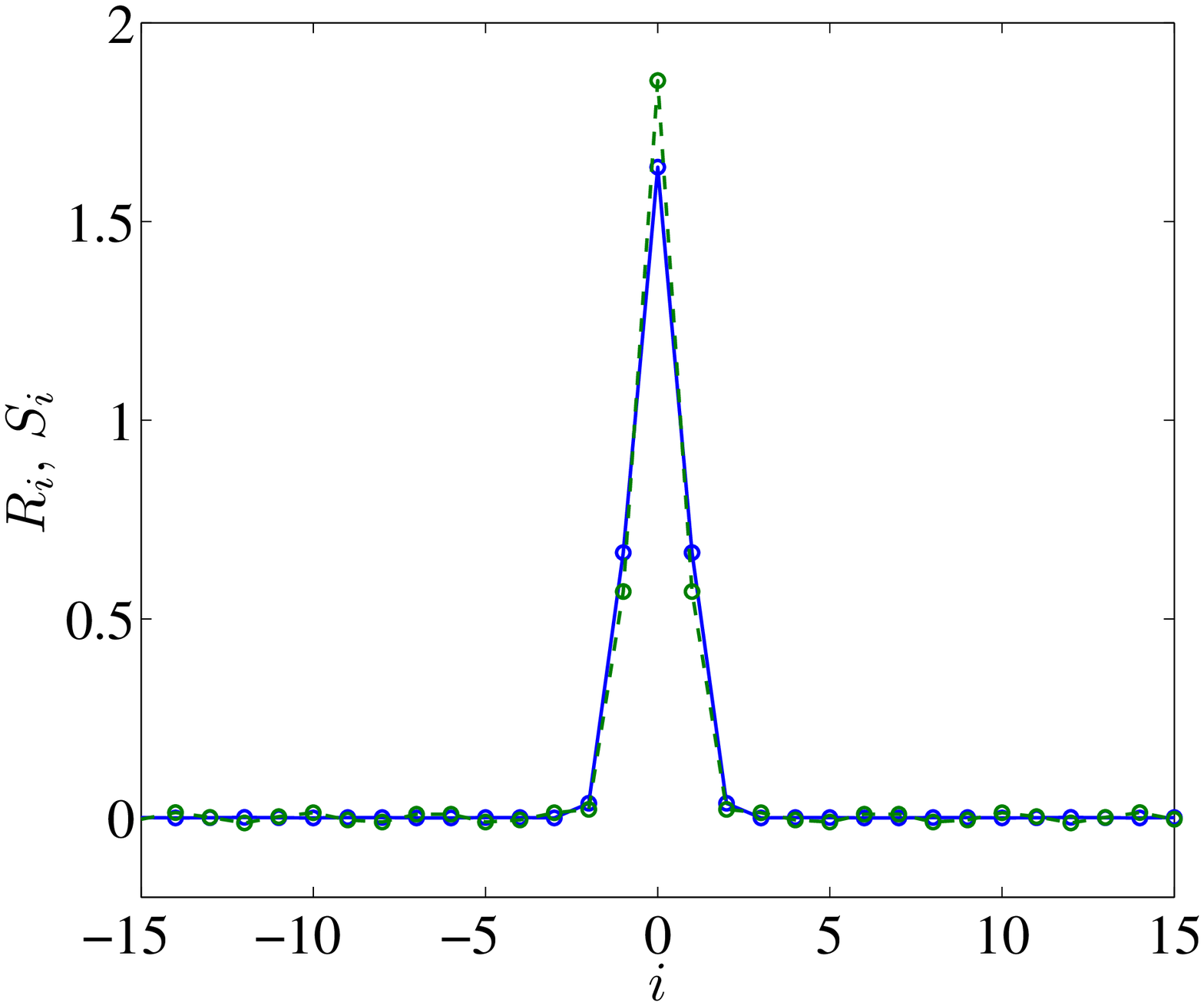}
}

\caption{The left panel shows the solution of
the continuum iterative processing of (\ref{eqn7}) and (\ref{eqn8})
in the co-traveling frame with $c=1$ for the fields $R$ and
$S$ as functions of $\xi$. The right panel identifies the
values of the field ordinates on the sites $i$ of the lattice. $R_i$ is
shown by the circles connected by the solid (blue) line, while $S_i$
by the circles connected by the dashed (green) line.}
\label{fig1}

\end{figure}

Subsequently, such nanoptera are ``released'' on the lattice
and are found to indeed propagate with the prescribed speed
of $c=1$, {\it without modifying their nanopteronic 
character}, as is clearly demonstrated for over 100
time (and space) units in Figure~\ref{fig2}.
The figure shows the two strain fields $\Delta_i(t)$ and $d_i(t)$ as
contour plots of space $i$ and time $t$ and the unit slope
propagation is clearly indicative of their robustness.
The initial and final profiles of  
the two fields are shown in the left insets (of the left
and right panel, respectively).
The right insets indicate the evolution of the maximum
of the field as a function of time. Both insets reveal
the unscathed traveling  nature of the resulting waveforms.
The tails are evident in the logarithmic scale evolution of
the strain fields shown in the bottom panels.

\begin{figure}[tbph]

\centerline{
\includegraphics[width=9.cm]{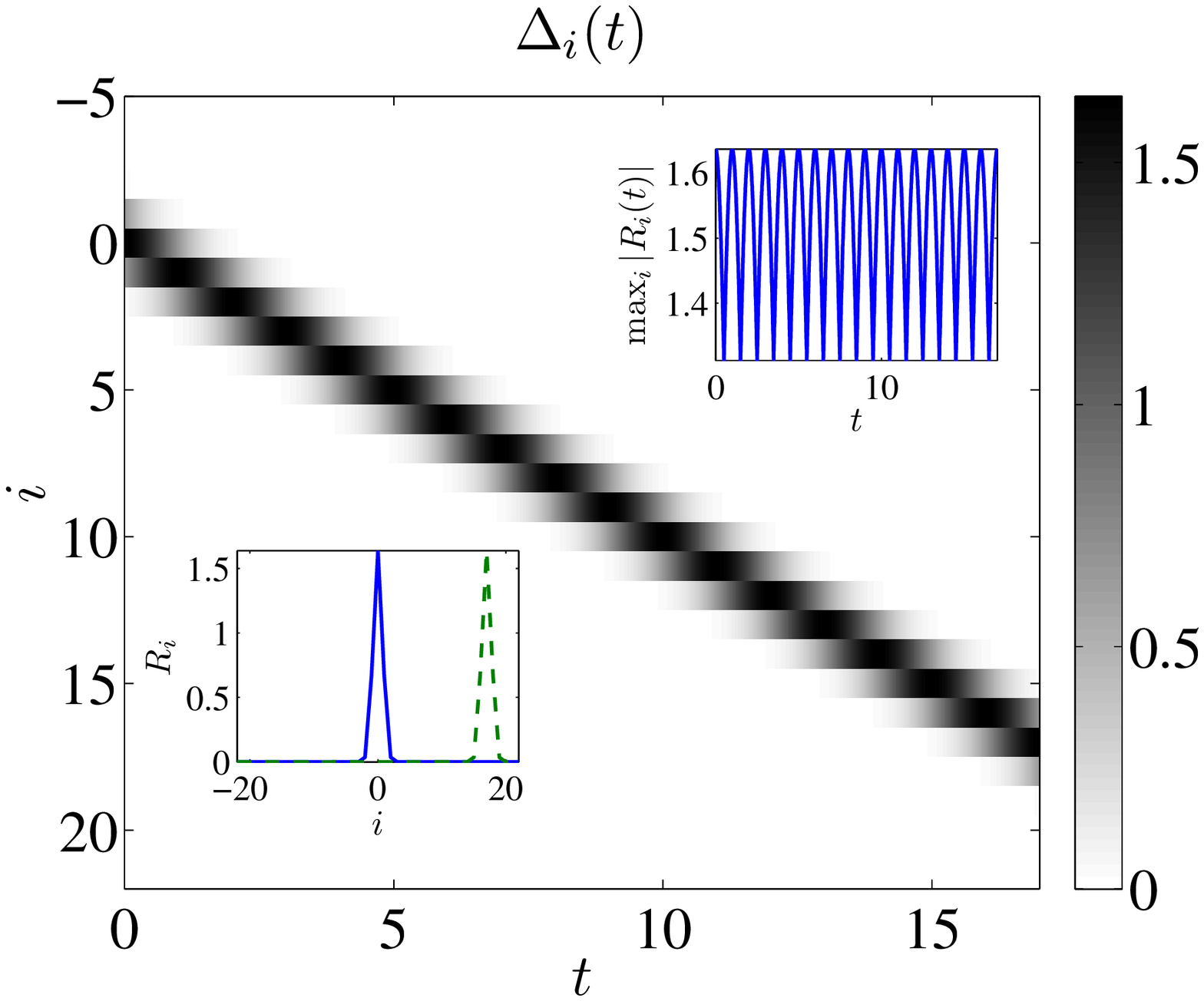}
\includegraphics[width=9.cm]{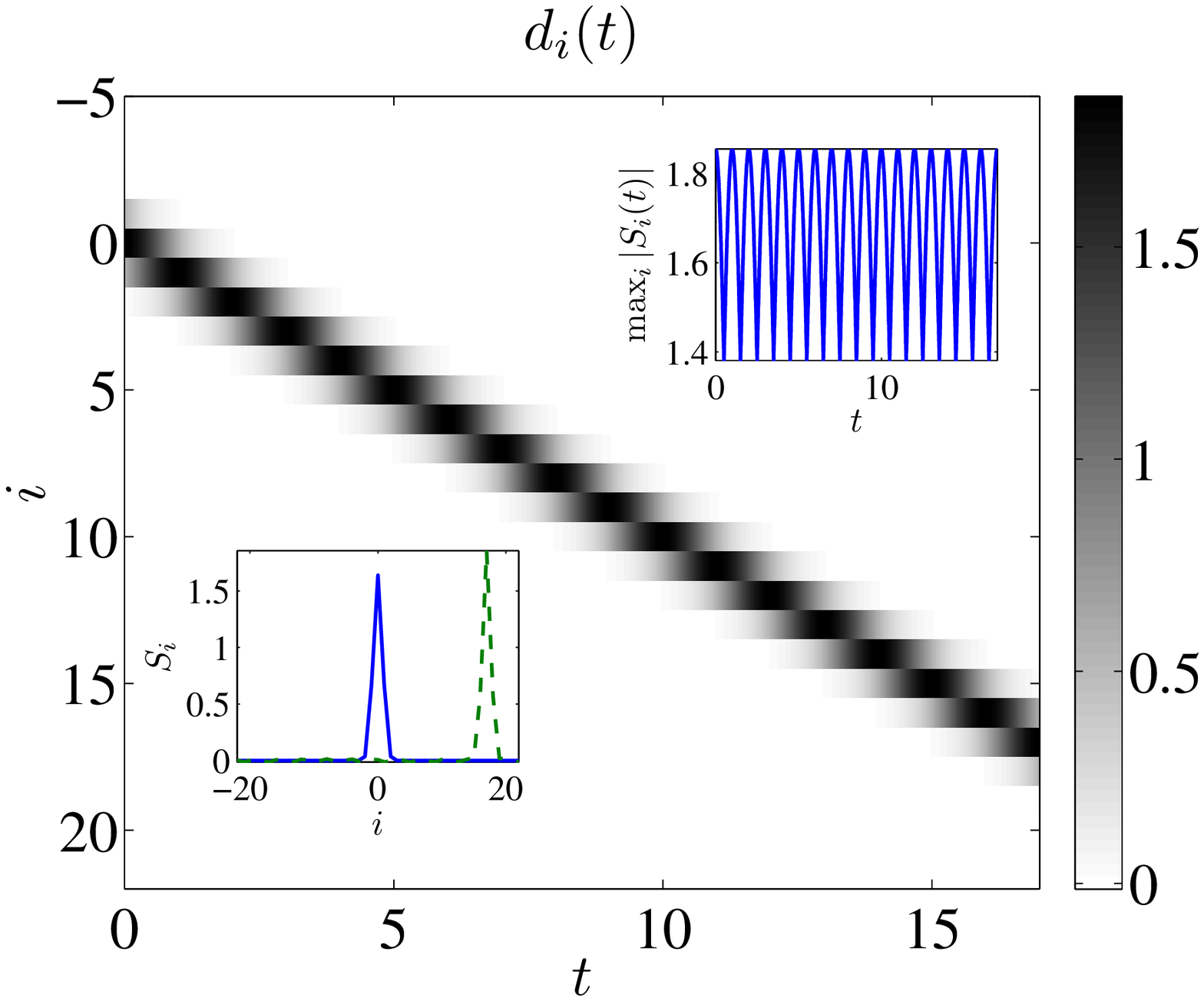}
}
\centerline{
\includegraphics[width=9.cm]{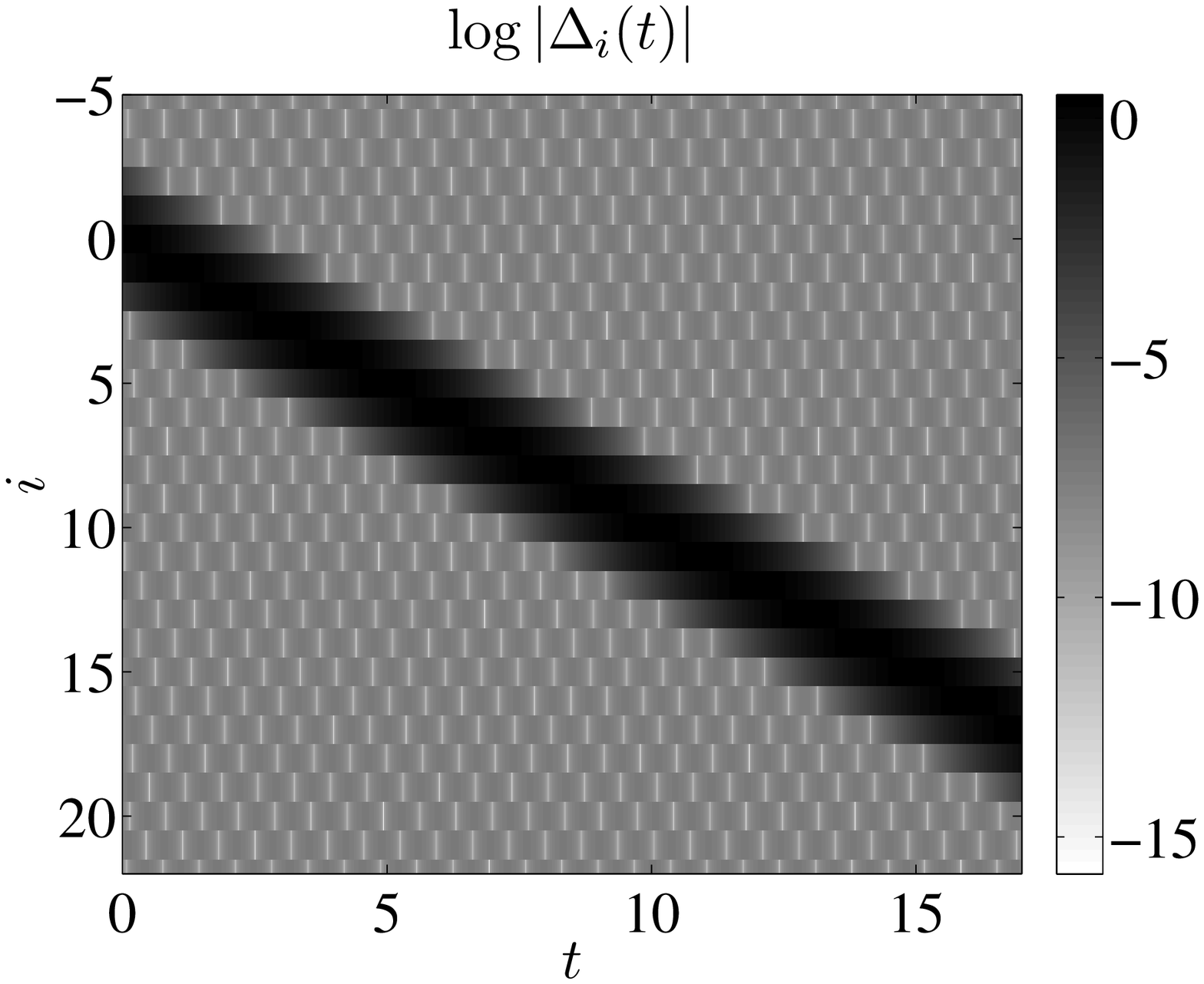}
\includegraphics[width=9.cm]{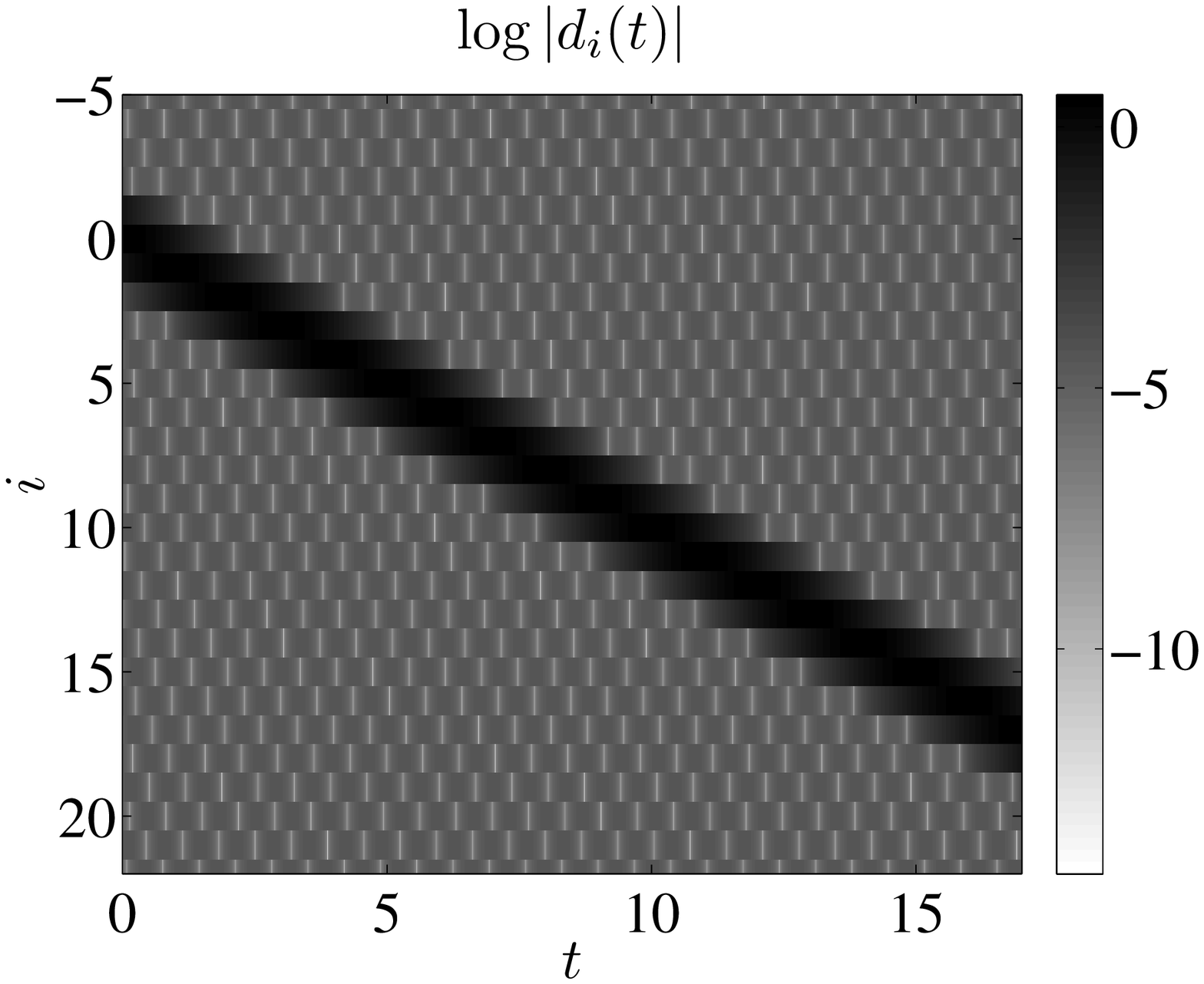}
}

\caption{The evolution of the two fields $\Delta_i(t)$
and $d_i(t)$ for nanopteron initial conditions 
in the case of $\nu=0.05$ and unit speed
(notice the propagation slope). The  right insets
show the time oscillation of the solution maximum,
while the left ones show the initial and final
solution configurations revealing their robust 
propagation. The bottom panels illustrate the logarithmic
scale of the relevant evolution, rendering clear the presence of the
tails.}
\label{fig2}
\end{figure}

\section{Conclusions and Future Challenges}

In the present work, we have studied the wave propagation in a novel granular
structure which consists of a chain of hollow beads
interacting via Hertzian contacts, and containing linear resonators.
We have seen that such a local oscillator provides a gateway
for the energy of the primary traveling pulses, well-known
to propagate unimpeded in the regular elastic chain, to be
gradually removed. This leads to a systematic decrease of the
pulse's amplitude. Adapting the recent method of~\cite{ref26}
to this complex, non-autonomous case, we were able
by means of ODE techniques (solving for the attachment
and back-substituting  in the shell equation, assuming
a decaying pulse in the shell equation and integrating
it between suitable instants in time) to capture the
pulse's decrease of amplitude. This was done by a 
nonlinear map approach that was found to be highly efficient,
under suitable assumptions. It is important to note, that the similar approach with modification has been applied in \cite{ref40}, and its results were found very efficient in depicting the evolution of the primary pulse amplitude.

Nevertheless, this amplitude and pulse decay were not the sole
features of this novel chain. Instead, such decay gave rise
to the emergence of secondary pulses that eventually overtake
the primary pulse, leading to a modulated waveform. Under suitable
conditions, this modulated waveform was found to self-organize
into a robust traveling wave that detached from the background
``radiation'' and was observed to propagate unhindered 
(and with constant amplitude and speed) through the lattice.
These waves were studied in more detail and were identfied as
weakly nonlocal solitary waves i.e., nanoptera. A formulation of
the existence problem of such waves was provided in Fourier (and real) 
space
which enabled the identification of their dominant wavenumber
$k_0$ and of their temporal frequency of tail oscillation 
$\omega=c k_0$, as well as its exact proof-of-principle computation. 
The frequency of tail oscillation was identified as the frequency
of the relative out-of-phase oscillation of the outer shell and the
inner attachment. 

Naturally, this study opens a number of new directions that are worthy
of examination in their own right. For instance, it would be
particularly interesting to explore in their own right the
nanopteronic solutions of the model from a rigorous mathematical
perspective and establish their regularity and stability properties,
as well as to develop robust numerical schemes for identifying
such coherent structures bearing a spatial tail.

Going beyond the present model, one of the critical
assumptions herein was that the oscillator connecting the outer shell
and inner attachment is linear. It would be interesting to explore the
possibility of soft or hard anharmonic oscillators and to examine
the fate of the nanoptera in the presence of such anharmonicity.
It is tempting to conjecture that they would persist on the basis
of the argument associated with the solitary wave tails, yet a systematic
study of the properties of such waves would certainly be warranted.

On the other hand, here we have restricted our considerations to the
``highly nonlinear'' limit (from the point of view of the outer shell 
Hertzian interactions). In that light, we have assumed that the shells
do not sustain an external, so-called pre-compression, force. Yet,
the existence of such precompression would provide a
linear limit to the problem that is of particular interest in its
own right in such mass-in-mass ``metamaterials''. The examination of
the linear and nonlinear properties of pre-compressed crystals would be
a topic for future studies in its own right.
A particularly relevant feature to explore in this context stems
from the comparison with~\cite{ref29}. There it was
found that the models with precompression spontaneously produce
nanoptera with oscillations on both sides, while in the absence
of precompression, only ones with extended waves on one side
arose. It would be interesting to explore a systematic
comparisons with these findings.

These themes are currently under investigation and will be reported in future 
works.

{\it Acknowledgements}. P.G.K is grateful to Ricardo Carretero and Haitao Xu for some technical assistance with some of the figures. He also gratefully acknowledges support from US-ARO under grant W911NF-15-1-0604 and from US-AFOSR under grant FA-9550-12-1-0332. G.T. acknowledges financial support from 
FP7-CIG (Project 618322 ComGranSol).

\section{Appendix}

It should be noted in passing here that a very accurate, analytically tractable Pad{\'e} approximation of the primary pulse
wave has been derived recently in ~\cite{ref27} in the form

\begin{equation}
\label{eq201}
\tilde{{S}}(\tau )\cong \left( {\frac{1}{q_{0} +q_{2} \tau^{2}+q_{4} \tau 
^{4}+q_{6} \tau^{6}+q_{8} \tau^{8}}} \right)^{2},
\end{equation}
where $\tilde{{S}}(\tau )$ denotes the normalized (with respect to a time 
shift) solitary wave solution of the homogeneous granular chain, 
$\tau $is a normalized time,$q_{0} ,q_{2} ,q_{4} ,q_{6} ,q_{8} $ are 
universal constants given explicitly in~\cite{ref27}. 

Unfortunately, this functional form of the solitary pulse is rather cumbersome and leads to the unnecessary complications
in the analysis devised in the present study. Thus, to overcome this difficulty in the present study the
approximation of the solitary wave profile is sought in the following functional form

\begin{equation}
\label{201}
\Delta_i^s(\tau)=S^{s}\left(\tau-i\right)=
\begin{cases}
B \cos^4(\alpha(\tau-i)), &  \tau\in \left[{i-\frac{\pi}{2\alpha} , i+\frac{\pi}{2\alpha} } \right] \\
0, & \text{otherwise}
\end{cases}
\end{equation}

The theoretical procedure of finding the fitting parameters $B$ and $\alpha$  comprises the following two rather simple stages. At the first stage we assign 
the amplitude to be equal to the amplitude of the particular, exact, solitary wave solution propagating with the unity phase shift ($B=1.4954$). This value can be easily deduced from either numerical simulations or semi-analytical 
techniques e.g. the English-Pego procedure \cite{english}. The second fitting parameter   is computed by equating the areas covered by the exact and the approximated solitary wave profiles, yielding the following simple algebraic relation.

\begin{equation}
\label{202}
\alpha = \frac{3\pi B}{8\int\limits_{-\infty }^{+\infty} {S_{\rm exact}\left(\tau \right)d\tau }} = 0.6446.
\end{equation}
In fact, the derived parameters $\alpha$ and $B$  can be considered as universal, as due to the homogeneity of the nonlinear lattice under investigation, the derived approximation can be properly scaled to fit to the arbitrary solitary wave profile, having the following form. 
\begin{equation}
\label{203}
S_{i}(\tau)=AS^{s}\left(A^{1/4}\tau-i\right)
\end{equation}
Here $A$ is the ratio between the amplitude of the arbitrary, solitary wave profile and the normalized one ($S^{s}\left(0\right)$). The comparison between the exact solution and the approximate solution is illustrated in Figure~\ref{fig200}.

We can see that this approximation is fairly adequate 
throughout the support of the relevant wave.
\begin{figure}[htbp]
\centerline{
\includegraphics[width=0.7\textwidth]{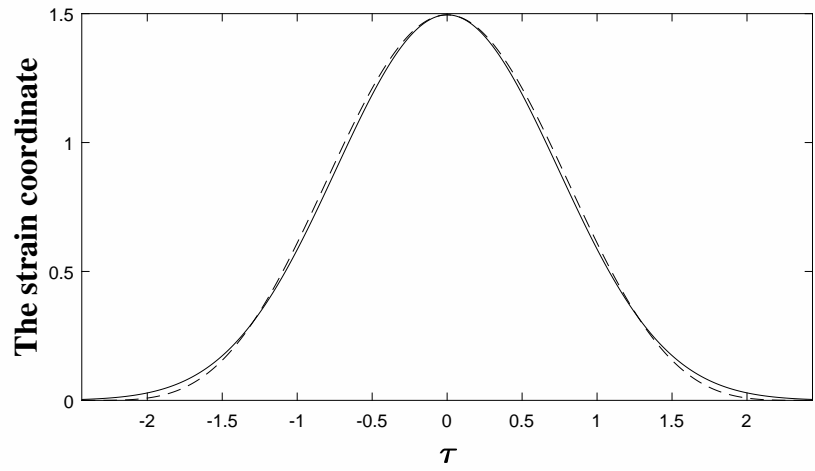}}
\caption{The comparison between the exact solution and the approximate one (\ref{201}). The solid line shows the exact solution and the approximate one is denoted by the dashed line.}
\label{fig200}
\end{figure}
\end{document}